\documentclass[11pt]{article}
 
\usepackage{amsmath,amsthm}
\usepackage{amsthm}
\usepackage{amssymb}
\usepackage{amsfonts} 
\usepackage{color}
\usepackage{hyperref}
\usepackage[latin1]{inputenc}
\usepackage{float}
\usepackage{graphicx}

\newtheorem{thm}{Theorem}[section]

\newtheorem{propn}[thm]{Proposition}

\newtheorem{rem}[thm]{Remark}

\newtheorem{defn}[thm]{Definition}

\newtheorem{cor}[thm]{Corollary}

\def\1{\'{\i}}

\begin{document}

\begin{center}
{\Large{\bf{Unified structures for solutions of Painlev\'e equation II and Somos-4 like relations for the tau functions.}}}

\bigskip 
\bigskip

\begin{center}
Federico Zullo$^{1,2}$, Maria Grazia Naso$^{1}$, Elena Vuk$^{1}$
\end{center}
\noindent
$^{1}$DICATAM, Universit\`a degli Studi di Brescia
via Branze, 38 - 25123 Brescia, Italy \\
$^{2}$INFN, Milano Bicocca,
Piazza della Scienza 3, Milano, 20126, Italy\\

\end{center}

\bigskip\bigskip

\begin{abstract}
\noindent We present certain general structures related to the solutions of Painlev\'e equation II and to the solutions of the differential equation satisfied by the corresponding Hamiltonian equations, together with the tau functions. By taking advantage of the B\"acklund transformations we find different explicit rational expressions linking the solutions of Painlev\'e equation II, Painlev\'e equation XXXIV and the Hamiltonians with the tau functions. Wronskians among different tau functions and the derivatives of the tau functions themselves will be expressed in terms of rational functions of tau functions too. A non-autonomous Somos-4 type relation solved by these functions is given. For the Somos-4 type relation we consider degenerate cases through the use of suitable parameters inserted into the equations: the autonomous case solvable in terms of Weierstrass elliptic functions, the case corresponding to the Yablonskii-Vorob'ev polynomials, the Airy-type solutions and the more general transcendental case. 

\end{abstract}

\bigskip\bigskip 

\noindent

\noindent
KEYWORDS: Painlev\'e II, Somos-4 equation,Weierstrass elliptic functions, Yablonskii-Vorob'ev polynomials, Airy functions, tau functions.

\section{Introduction}
The Painlev\'e II equation is usually written in the following form:
\begin{equation}\label{p2}
\frac{d^2 y}{dz^2}=2y^3+zy+\alpha,
\end{equation}
where $\alpha$ is a parameter. It is well known that this equation possesses, for integer and half-integer values of $\alpha$, rational solution and solution explicitly expressible in terms of Airy functions. All other solutions are transcendental. One of the differences among these solutions is in the distribution of the zeros in the complex plane: they are finite in number for rational solutions, are infinitely many in the other two cases but the order of growth of the solutions are different: it is $3/2$ for the Airy type solutions and $3$ for the other.  If instead of the distribution  of the zeros one takes into account, for example, the asymptotic behaviour of the solutions of (\ref{p2}) on the real axis, then another variety of solutions with different analytical properties are obtained also in the case of a fixed value of the parameter $\alpha$ (see e.g. \cite{C1} and references therein).  This  wide variety of type of solutions prevented the development of a unified notation for  the Painlev\'e transcendents. A unifying trait for the solutions of (\ref{p2}) is that the poles are all simple with residues plus or minus 1. Indeed, all the solutions of equation (\ref{p2}) can be represented as the difference of two functions having simple poles with residues $+1$: these functions, usually called $\sigma$ functions \cite{Clarkson}, from a mechanical point of view are Hamiltonian functions and for this reason are also symbolized with an $H$ (see e.g. \cite{Okamoto1, Okamoto2}). The Painlev\'e equations are indeed Hamiltonian, i.e. they possess an Hamiltonian structure; the relative Hamilton equations correspond to the Painlev\'e equations. It is the Hamiltonian, as a function of $z$, that indeed appears more frequently in the applications. 

In this paper, following \cite{HZ2}, we introduce some parameters in the Hamilton equations also to elucidate the connections between the Painlev\'e equation II and  a degenerate case involving the Weierstrass elliptic functions.  Further details on why we find useful to introduce the parameters will be given also in the next Sections. Now it is important and significant to observe that the link with the Weierstrass elliptic functions turns out to be not an accident but it reflects some general structures of the solutions of the Painlev\'e equation II and of the corresponding tau functions. These general structures, that we are going to report and discuss, represent the main finding of the work. Actually, these structures reflect the common Hamiltonian structure and the common existence of canonical transformations, i.e. B\"acklund transformations, among different solutions of the Painlev\'e equation (apart the elliptic case when one has auto-B\"acklund transformations). So we are going to consider the chain of solutions of the Painlev\'e equation II, together with the chain of solutions of the Painlev\'e XXXIV equation and the solutions of the differential equation satisfied by the Hamiltonian functions. The seed solution can be any of the possible type of solutions that the equations admit on: transcendental, Airy-type solutions, rational functions (or polynomials for the tau functions) or elliptic function type solutions. Differential equations, difference equations and differential-difference equations for the chains of solutions will be given. Some of them are well-known, other are new. More importantly we will show that all these functions (i.e. functions solutions of Painlev\' e II equation, Painlev\'e XXXIV equation, the equation for the Hamiltonian function and the tau functions) and all their type of solutions  can be expressed as \emph{rational functions} of different tau functions. The Wronskians between different tau functions and the derivatives of the tau functions themselves share this property too. Further, we will show that the tau functions solve a non-autonomous Somos-4 relation that, in the degenerate Weierstrass case, becomes the well-known Somos-4 relation giving integer sequences (see \cite{H} or \cite{Swart}). Essentially, the tau functions are the bricks to build a generalized structure related to the solutions of the Painlev\'e II equation.

The paper is organized as follows: in  Section \ref{sec2} we introduce a set of parametric Hamiltonians together with the corresponding canonical variables related to the Painlev\'e equation (\ref{p2}). Also, we introduce the B\"acklund transformations, essentially as given by Okamoto \cite{Okamoto2}. 

In Section \ref{sec3}  we get a bilinear equation satisfied by the Hamiltonian functions and then two explicit third order recurrences. The compatibility between these two gives a second order quadratic recursion. This second order equation can be reduced, thanks to the explicit link between the Hamiltonians and the solutions of the Painlev\'e II equation, to the well known discrete Painlev\'e I in alternative form (see e.g. equation 1.21 in \cite{FGR}). 

In Section \ref{sec4} we will show how all the solutions of the main differential equation of this work, i.e. equation (\ref{secor}), can be formally reduced to the solutions of Riccati equations whose coefficients are certain combinations of Hamiltonians themselves (hence transcendental functions in the general case). By introducing the tau functions corresponding to the Hamiltonians one gets four different bi-linear equations for the tau functions. Further, we will show that the B\''acklund transformations introduced essentially reduces to a quadratic expression for the Wronskian between two successive tau functions. Most of the formulae up to this Section are preparatory and are well-known in literature. When will be given results that, as far as we know, represent novelties in the literature, will be explicitly underlined. 

In Section \ref{sec5} we will show that the bi-linear equations, together with the Toda equation, lead to two explicit linear equations for the Wronskian between two successive tau functions. This gives the possibility to express the derivative of the tau functions as rational functions of adjacent tau functions. Further, we get explicit expressions for the solutions of the Painlev\'e equation II, Painlev\'e equation XXXIV and the Hamiltonians in terms of rational combinations of tau functions. Also, a fifth order explicit recursion for the tau functions, firstly given in \cite{JM}, will be given. 

In Section \ref{sec6} we analyze the divisibility properties of the tau functions (or rational combination of them) in the ring of entire functions. These divisibility properties will bring forth a Somos-4 like relations solved by the tau functions. In this relation it enters another entire function, that we denote with $y_n$, that seems to share some properties in common with the tau functions themselves. Indeed, we are able to show that, in the degenerate Weierstrass case, $y_n$ reduces to $\tau_n$. We will show certain analytical and numerical characteristics of $y_n$ also in the polynomial case and in the Airy case.

Finally, in the Conclusions, after reviewing the main finding of the work, we will try to underline the possible usefulness of the new formulae, for example in the possibility to specify properties of the tau functions or of the solutions of Painlev\'e XXXIV equation valid throughout the complex plane. Also, we will specify some possible extension of the research that, according to our point of view, seems to be relevant. 

\section{Canonical transformations}\label{sec2}
We consider the following set of Hamiltonian functions, labelled by the discrete index $n \in \mathbb{Z}$, written in terms of the canonical variable $(q_n,p_n)$ as: 
\begin{equation}\label{HamH}
H_n=\frac{p_n^2}{2m}-ap_n(q_n^2+bz+c)-e_n q_n -\frac{mg_2}{2a^2}
\end{equation}
The last term, proportional to $g_2$, is just a constant and it could be removed by setting it equal to zero, but we will keep it since it will be useful in the rest of the paper. The other constants are $m, a, b, c, e_n$, with $e_n=e+2abmn$, $a$ and $m$ in general different from zero. Apart $n$, all the constant can be complex in principle. The previous set of Hamiltonians is essentially the one given by Okamoto \cite{Okamoto1}-\cite{Okamoto2}. The corresponding Hamilton equations are given by
\begin{equation}\label{mot}\begin{split}
&\frac{dq_n}{dz}=\frac{p_n}{m}-a(q_n^2+bz+c),\\
& \frac{dp_n}{dz}=e_n+2ap_nq_n.
\end{split}\end{equation}
The differential equations satisfied by $q_n$ and $p_n$ are respectively given by
\begin{equation}\label{eqmot}\begin{split}
&\frac{d^2q_n}{dz^2}=2a^2q_n^3+2a^2\left(bz+c\right)q_n+\frac{e_n}{m}-ab,\\
&p_n\frac{d^2p_n}{dz^2}=\frac{1}{2}\left(\frac{dp_n}{dz}\right)^2+\frac{2a}{m}p_n^3-2a^2(bz+c)p_n^2-\frac{e_n^2}{2}.
\end{split}\end{equation} 
The first equation is the Painlev\'e II, whereas the second one is known as  Painlev\'e XXXIV (in the Ince's notation \cite{Ince}). It is clear that we could get rid of various constants, for example $a$ and $m$, just by rescaling the dependent and independent variables. In the applications however, for example in the scaling reductions of partial differential equations, usually the independent variable is a space or time variables or a precise combination of both and through the scaling one obtains equation (\ref{p2}) only after other reductions. This is one of the reasons we find useful to leave the parameters in equations (\ref{HamH}) and (\ref{mot}).  

Due to the canonical structure of the equations of motion, the derivative of $H_n$ as a function of $z$ is given by
\begin{equation}\label{dh}
\frac{dH_n}{dz}=\frac{\partial H_n}{\partial z}=-abp_n
\end{equation}
and from (\ref{mot}) it follows that $H_n$ solves the following second order differential equation
\begin{equation}\label{secor1}
\left(\frac{d^2H_n}{dz^2}\right)^2+\frac{2}{bm}\left(\frac{ dH_n(z)}{dz}\right)^3+4a^2c\left(\frac{dH_n}{dz}\right)^2+4a^2b\frac{dH_n}{dz}\left(z\frac{dH_n}{dz}-H_n\right)-2bmg_2\frac{ dH_n(z)}{dz} - a^2b^2e_n^2=0.
\end{equation}

Equation (\ref{secor1}) possesses meromorphic solutions \cite{GLS}. Indeed, by the Painlev\'e test, it is not difficult to see that the poles are simple with  residues equal to $2bm$, so we let $H_n(z)=2bmh_n(z)$ with $h_n(z)$ having all poles with residue 1. The corresponding differential equation for $h_n$ reads
\begin{equation}\label{secor} 
(\ddot{h}_n)^2 + 4 (\dot{h}_n)^3+4a^2(bz+c)(\dot{h}_n)^2-(4a^2bh_n+g_2)\dot{h}_n-\frac{a^2 e_n^2}{4m^2}=0.
\end{equation}
This is our main differential equation. It possesses rational solutions, solutions that can be write in terms of elliptic functions (and the corresponding degenerate case given in terms of trigonometric functions), solutions written in terms of Airy functions and their derivatives and transcendental solutions with  order of growth $\leq 3$. We are not aware of any proof that the transcendental solutions of (\ref{secor1}), apart the rational, elliptic and the Airy type solutions, have order of growth exactly 3 when $b \neq 0$. When $b=0$ all the solutions have order of growth 2 since the general solution can be reduced to Weierstrass elliptic functions (indeed, in this case it corresponds to the equation for the Weierstrass $\zeta$ function). As we shall see there is a further degenerate case corresponding to solutions given in terms of trigonometric function, but the order of growth will remain equal to 2. When $b \neq 0$ a simple shift will eliminate the dependence of $h_n$ on $g_2$. However, as explained above, we want to keep the constant $g_2$ in (\ref{secor}): it corresponds to the elliptic invariants of the Weierstrass functions when $b=0$. To further elucidate the connection between the Weierstrass elliptic functions and equations (\ref{eqmot}), we make a further observation. If we introduce the function $P_n=\frac{a}{2m}p_n$, the differential equation for $p_n$ in (\ref{eqmot}) becomes
\begin{equation}
P_n\ddot{P_n}=\frac{1}{2}\dot{P}_n^2+4P_n^3-2a^2(bz+c)-\frac{e_n^2a^2}{4m^2}.
\end{equation}
This equation, by taking into account the relation $P_n=-\dot{h}_n$ (see equation (\ref{dh})), can be equivalently written as
\begin{equation}
\left(\dot{P}_n\right)^2=4P_n^3-g_2P_n+\frac{e_n^2a^2}{4m^2}-4a^2\left((bz+c)P_n^2+bP_nh_n\right),
\end{equation}
and it reduces to the equation for the Weierstrass $\wp$ function when $b=c=0$ (also for $b=0$ and $c\neq 0$ through a simple shift of the dependent variable).

Notwithstanding, for $b\neq 0$, it could be useful to introduce the function $k_n$ related to $h_n$ by
\begin{equation}\label{kn}
h_n(z)=-\frac{g_2}{4a^2b}+k_n(z).
\end{equation}

Now let us now consider the following set of canonical transformations (also introduced by Okamoto \cite{Okamoto2}):
\begin{equation}\label{tra}\begin{split}
&q_{n+1}=-q_n-\frac{e_n}{ap_n},\\
&p_{n+1}=-p_n+2amq_{n+1}^2+2am(bz+c).
\end{split}\end{equation}
The function $H_{n}(p_n,q_n,z)$, and hence $h_{n}(p_n,q_n,z)$, is preserved by the transformations, i.e. 
$$H_{n}(p_n(p_{n+1},q_{n+1}),q_n(p_{n+1},q_{n+1}),z)=H_n(p_{n+1},q_{n+1},z).$$
Also, the generating function of the non-autonomous canonical transformations (\ref{tra}) is given by
\begin{equation}
F_n=-\frac{e_n}{a}\ln(q_n+q_{n+1})-\frac{2am}{3}q_{n+1}^3-2ma\left(c+bz\right)q_{n+1}.
\end{equation}
Since the Hamiltonian system defined by the Hamiltonian (\ref{HamH}) is non autonomous, it follows that $(q_{n+1},p_{n+1})$, as functions of $z$, are the integral curves of the new Hamiltonian \cite{Z0}
\begin{equation}\label{HHq}
H_n(p_{n+1},q_{n+1},z)+\frac{\partial F}{\partial z}=H_n(p_{n+1},q_{n+1},z)-2mab q_{n+1}=H_{n+1}(p_{n+1},q_{n+1},z).
\end{equation}
Now we are going to refer all to the function $h_n$ and not to $H_n$. From (\ref{HHq}) it follows that the function $q_{n+1}$ is proportional to the difference $h_n-h_{n+1}$:
\begin{equation}\label{Hnn}
h_{n}-h_{n+1}=aq_{n+1}.
\end{equation}
The equations of motion (\ref{mot}) and the canonical transformations (\ref{tra}) give:
\begin{equation}
\frac{\dot{p}_n}{p_n}=\frac{e_n}{p_n}+2aq_n=a(q_n-q_{n+1}).
\end{equation}
The previous, together with (\ref{tra}), gives
\begin{equation}\label{qnn}\begin{split}
&a(q_n-q_{n+1})=\frac{\dot{p}_n}{p_n}=\frac{\ddot{h}_n}{\dot{h}_n},\\
&a(q_{n}+q_{n+1})=-\frac{e_n}{p_n}=\frac{a e_n}{2m\dot{h}_n}.
\end{split}\end{equation}
Equations (\ref{Hnn}) and (\ref{qnn}) are still valid if one substitutes $h_n$ with $k_n$ given in equation (\ref{kn}). From equations (\ref{qnn}) it follows:
\begin{align}
aq_n=\frac{a e_n}{4m\dot{h}_n} +\frac{\ddot{h}_n}{2\dot{h}_n}, \label{qnnp1a}\\
aq_{n+1}=\frac{a e_n}{4m\dot{h}_n} -\frac{\ddot{h}_n}{2\dot{h}_n} \label{qnnp1b},
\end{align}
and by evaluating the first equation in $n \to n+1$ and subtracting it yields:
 \begin{equation}
2m\left(\ddot{h}_{n+1}\dot{h}_{n}+\dot{h}_{n+1}\ddot{h}_{n+1}\right)+a\left(e_{n+1}\dot{h}_{n}-e_n\dot{h}_{n+1}\right)=0.
\end{equation}
The previous can  be integrated once, giving:
\begin{equation}\label{dotHn}
2m\dot{h}_{n+1}\dot{h}_{n}+a\left(e_{n+1}h_n-e_n h_{n+1}\right)=C_1,
\end{equation}
where, as we will now show, the constant on the right hand side must be equal to $-mg_2/2$. Indeed, equations (\ref{qnn}) implies
\begin{equation}\label{aq}
aq_{n+1}=-\frac{e_n+\dot{p}_n}{2p_n},
\end{equation}
whereas from (\ref{Hnn}) and (\ref{aq}) one gets \cite{Okamoto2}
\begin{equation}\label{Hn1}
h_{n+1}=h_n+\frac{2m\ddot{h}_n-ae_n}{4m\dot{h}_n}.
\end{equation}
Equation (\ref{Hn1}) can be considered the B\"acklund transformation for the Hamiltonian $h_n$: if $h_n$ is a solution of (\ref{secor}), then $h_{n+1}$ given by (\ref{Hn1}) is a solution of (\ref{secor}) evaluated in $n \to n+1$. This can be checked also by inserting equation (\ref{Hn1}) in the equation  (\ref{secor}) evaluated in $n \to n+1$ and considering equation (\ref{secor}) itself and its differential consequences. Now we can insert $h_{n+1}$ from (\ref{Hn1}) into (\ref{dotHn}): we get a third order equation for $h_n$ and the compatibility with (\ref{secor}) implies $C_1=-mg_2/2$. We also notice that equation (\ref{dotHn}) written in terms of the shifted hamiltonians $k_n$ (\ref{kn}) is
\begin{equation}
2m\dot{k}_{n+1}\dot{k}_{n}+a\left(e_{n+1}k_n-e_n k_{n+1}\right)=0.
\end{equation}

Now it is possible to get the corresponding inverse transformation of (\ref{Hn1}) by looking at the inverse transformations of (\ref{tra}), given by
\begin{equation}\label{invtra}\begin{split}
&q_{n+1}=-q_n-\frac{e_n}{ap_{n+1}},\\
&p_{n+1}=-p_n+2amq_{n}^2+2am(bz+c).
\end{split}\end{equation}
By repeating the previous discussion to the inverse transformations (\ref{tra}) we can write the inverse of the B\"acklund transformations for $h_n$ (\ref{Hn1}) as:
\begin{equation}\label{invHn1}
h_n=h_{n+1}+\frac{2m\ddot{h}_{n+1}+ae_{n+1}}{4m\dot{h}_{n+1}}.
\end{equation}
We notice that this equation can be derived also from (\ref{qnnp1a}): by evaluating it in $n \to n+1$ and using (\ref{Hnn}) one gets (\ref{invHn1}). The transformation (\ref{Hn1}) gives a solution of the equation (\ref{secor}) evaluated in $n \to n+1$ from a solution of equation (\ref{secor}) itself. The transformation (\ref{invHn1}) pull back this solution. These two transformations define a discrete dynamics over $n$: in the next Section we will show how both (\ref{Hn1}) and (\ref{invHn1}) provide  certain recurrences for the functions $h_n$ and the corresponding tau functions.

\section{Recursions for the Hamiltonian functions}\label{sec3}
It is possible to get another bilinear equation involving the Hamiltonian functions $h_n$ and its derivatives. Adding (\ref{Hn1}) and (\ref{invHn1}) we get
\begin{equation}
 (h_{n+1}-h_n)(\dot{h}_{n+1}-\dot{h}_n) +\frac{\ddot{h}_{n+1}}{2}+\frac{\ddot{h}_n}{2}+\frac{a}{4m}(e_{n+1}-e_n) =0,
\end{equation}
and, by taking into account also that $e_{n+1}-e_n=2abm$ and integrating once
\begin{equation}\label{C2}
 (h_{n+1}-h_{n})^2+\dot{h}_{n+1}+\dot{h}_n +a^2bz=C_2.
\end{equation}
By inserting (\ref{Hn1}) in (\ref{C2}) and using (\ref{secor}) we see that the constant $C_2$ must be equal to $-a^2c$. So we get
\begin{equation}\label{linear}
 (h_{n+1}-h_{n})^2+\dot{h}_{n+1}+\dot{h}_n +a^2(bz+c)=0.
\end{equation}
The previous equation can be seen as a Riccati equation for $h_{n+1}$ or for $h_n$: the coefficients of the Riccati equation involve the adjacent Hamiltonian. We will return on this equation in Section \ref{sec3}. Now we will obtain an explicit expression for the derivative of $h_n$ in terms of Hamiltonian functions: this will give two explicit recursions for $h_n$ from equations (\ref{linear}) and (\ref{dotHn}).

By evaluating (\ref{invHn1}) in $n \to n-1$ and subtracting with (\ref{Hn1}) we get the following explicit expression for $\dot{h}_{n}$:
\begin{equation}\label{dhn}
\dot{h}_n=-\frac{ae_n}{2m(h_{n+1}-h_{n-1})}.
\end{equation}
It is possible to use equation (\ref{dhn}), together with (\ref{linear}) and (\ref{dotHn}) to obtain an explicit recursion for the function $h_{n+3}(z)$ in terms of $h_{n+2}(z)$, $h_{n+1}(z)$ and $h_{n}(z)$. From (\ref{linear}) we have
\begin{equation}\label{rec1}
 (h_{n+2}-h_{n+1})^2-\frac{ae_{n+2}}{2m(h_{n+3}-h_{n+1})}-\frac{ae_{n+1}}{2m(h_{n+2}-h_{n})} +a^2(bz+c)=0,
\end{equation}
that is, by isolating $h_{n+3}$:
\begin{equation}\label{h31}
h_{n+3}=h_{n+1}+\frac{ae_{n+2}}{2m}\frac{h_{n+2}-h_n}{(h_{n+2}-h_n)\left(a^2(bz+c)+(h_{n+2}-h_{n+1})^2\right)-\frac{ae_{n+1}}{2m}}.
\end{equation}
Equation (\ref{dotHn}) gives
\begin{equation}\label{h3in}
\frac{ae_{n+2}e_{n+1}}{2m}+(h_{n+3}-h_{n+1})(h_{n+2}-h_{n})(h_{n+1}e_{n+2}-h_{n+2}e_{n+1})+\frac{mg_2}{2a}(h_{n+3}-h_{n+1})(h_{n+2}-h_{n})=0,
\end{equation}
and, by isolating again $h_{n+3}$ we get:
\begin{equation}\label{h32}
h_{n+3}=h_{n+1}+\frac{a^2e_{n+1}e_{n+2}}{m(h_{n+2}-h_{n})(mg_2+2a(h_{n+1}e_{n+2}-h_{n+2}e_{n+1}))}.
\end{equation}
The compatibility between equations (\ref{rec1}) and (\ref{h31}) gives the following second order equation for $h_{n+2}$:
\begin{equation}\label{discr}
\begin{split}
&(h_{n+2}-h_{n+1})(h_{n+2}-h_n)(h_{n+1}-h_n)-\frac{m}{2ae_{n+1}}(h_{n+2}-h_{n})^2\left(4a^2bh_{n+1}+g_2\right)+\\
&-a^2(bz+c)\left(h_{n+2}-h_n\right)+\frac{ae_{n+1}}{2m}=0.
\end{split}
\end{equation}
Let us make some comments on the above recurrences. The dependence on $g_2$ disappears (as it must be) from (\ref{h3in}) and (\ref{h32}) if we make the shift (\ref{kn}). One has
\begin{equation}\label{h3in1}
\frac{ae_{n+2}e_{n+1}}{2m}+(k_{n+3}-k_{n+1})(k_{n+2}-k_{n})(k_{n+1}e_{n+2}-k_{n+2}e_{n+1})=0,
\end{equation}
and
\begin{equation}\label{h32kn}
k_{n+3}=k_{n+1}+\frac{ae_{n+1}e_{n+2}}{2m}\frac{1}{(k_{n+2}-k_{n})(k_{n+1}e_{n+2}-k_{n+2}e_{n+1})}.
\end{equation}
Equations (\ref{rec1}) and (\ref{h31}) remain unaltered by the shift instead. The recursion (\ref{discr}) becomes 
\begin{equation}\label{discrkn}
\begin{split}
&(k_{n+2}-k_{n+1})(k_{n+2}-k_n)(k_{n+1}-k_n)-\frac{2abm}{e_{n+1}}(k_{n+2}-k_{n})^2k_{n+1}+\\
&-a^2(bz+c)\left(k_{n+2}-k_n\right)+\frac{ae_{n+1}}{2m}=0.
\end{split}
\end{equation}

Equation (\ref{h31}) is related to the well-known discrete Painlev\'e I equation in alternative form (see e.g. \cite{FGR}, equation 1.21). Indeed, from equation (\ref{Hnn}) one has that $q_n$ solves the recursion 
\begin{equation}
\frac{e_{n-1}}{a(q_n+q_{n-1})}+\frac{e_n}{a(q_{n+1}+q_{n})}+2am(q_n^2+(bz+c))=0.
\end{equation}
This recursion for $q_n$ can be found more easily directly from the canonical transformations (\ref{tra}). The same transformations, together with their inverse, can be used to find a recursion involving only the $p_n$ functions. One gets (see also \cite{Conte}):
\begin{equation}\label{precursion}
\frac{e_n}{2ap_n}+\frac{p_{n+1}(p_{n+2}-p_{n})}{4me_{n+1}}=\frac{e_{n+1}}{2ap_{n+1}}-\frac{p_{n}(p_{n+1}-p_{n-1})}{4me_{n}}.
\end{equation} 
The previous is a third order difference equation for $p_n$. It is also possible to get a second order, second degree equation for these functions. Indeed, (\ref{tra}) entails
\begin{equation}\label{qnqn1}
q_n=\frac{p_n(p_{n+1}-p_{n-1})}{4me_n}-\frac{e_n}{2ap_n},\quad q_{n+1}=-\frac{p_n(p_{n+1}-p_{n-1})}{4me_n}-\frac{e_n}{2ap_n},
\end{equation}
and by inserting the expression for $q_{n+1}$ in (\ref{tra}) one has
\begin{equation}
\frac{p_n+p_{n+1}}{2am}-\left(\frac{ap_{n}^2(p_{n+1}-p_{n-1})+2me_{n+1}^2}{4ame_np_{n}}\right)^2=bz+c.
\end{equation}
It is also possible to get an explicit expression for $p_n$ in terms of the preceding functions $p_0,\ldots, p_n$. Indeed from (\ref{qnqn1}) we get for $n>1$
\begin{equation}\label{pexp}
\frac{p_n(p_{n+1}-p_{n-1})}{4me_n}=\frac{p_1(p_{2}-p_{0})}{4me_1}+\frac{e_n}{2ap_n}+(-1)^n\frac{e_1}{2ap_1}+\frac{1}{a}\sum_{k=1}^{n-1}(-1)^{n-k}\frac{e_k}{p_k}.
\end{equation}
It is interesting to notice that formulae (\ref{qnqn1}) and (\ref{pexp}) do not depend explicitly on $z$.

The other two recursions for $h_n$, i.e. (\ref{h32}) and (\ref{discr}), give an explicit expression for the Hamiltonian $h_n$ in terms of the $q_n$'s functions. From (\ref{h32}) one has
\begin{equation}
4a^2bh_n+g_2+\frac{2}{m^2}\left(e_{n+1}q_{n+2}-2abmq_{n+1}\right)+\frac{e_{n+1}e_{n+2}}{m^2(q_{n+1}+q_{n+2})(q_{n+2}+q_{n+3})}=0,
\end{equation}
whereas (\ref{discr}) or (\ref{discrkn}) give
\begin{equation}
4a^2bh_n+g_2=\frac{e_{n+1}^2}{m^2(q_{n+1}+q_{n+2})^2}+\frac{2a^2e_{n+1}(q_{n+2}^2+bz+c)}{m(q_{n+1}+q_{n+2})}+\frac{2a^2(2abmq_{n+1}-e_{n+1}q_{n+2})}{m}.
\end{equation}

Another comment: the difference equation (\ref{discr}), or (\ref{discrkn}), is quadratic in the three variables $h_{n}$, $h_{n+1}$ and $h_{n+2}$ (or $k_{n}$, $k_{n+1}$ and $k_{n+2}$). It is known that the B\"acklund transformations (\ref{tra}), and hence (\ref{discr}), produce rational solutions from rational solutions and special solutions written as rational combinations of Airy functions from the Airy-type solutions. This means that, at least in these cases, the discriminant of equation (\ref{discr}) with respect to one of the variables, say $h_{n+2}$, is a perfect square. Actually we will now show that this discriminant is always a perfect square and this is very useful in identifying the solutions of equation (\ref{discr}) and hence the explicit recursion for $h_{n+2}$ in terms of the other variables. Clearly this should be consistent with equations (\ref{h31}) and (\ref{h32}). At first sight it seems that (\ref{discr}) is different from (\ref{h31}) and (\ref{h32}), in the sense that (\ref{discr}) is a second order difference equation, whereas (\ref{h31}) and (\ref{h32}) are third order equations. Actually we will show that a solution of (\ref{discr}) is related to $h_{n-1}$. The main reason for this is the expression (\ref{dhn}) relating the derivative of $h_n$ to $h_{n-1}$. 

The discriminant $\Delta_n$ of equation (\ref{discr}) with respect to $h_{n+2}$ is
\begin{equation}\label{D1}
\Delta_n=\left((h_{n+1}-h_n)^2+a^2(bz+c)\right)^2+4a^2bh_{n+1}-\frac{2ae_{n+1}}{m}\left(h_{n+1}-h_n\right)+g_2.
\end{equation}  
In order to identify if $\Delta_n$ can be written as a perfect square we use the identity (\ref{Hn1}) for $h_{n+1}$ in terms of $h_n$. By carefully looking at the corresponding equation for $\Delta_n$ we arrive at the following expression:
\begin{equation}\label{D2}
\Delta_n=\left(\frac{\left(2m\ddot{h}_n-ae_n\right)^2+16a^2m^2(bz+c)\dot{h}_n^2+32m^2\dot{h}_n^3}{16m^2\dot{h}_n^2}\right)^2.
\end{equation}
Indeed, the difference between (\ref{D1}) and (\ref{D2}) is proportional to the left hand side of equation (\ref{secor}) and therefore is zero. By taking into account the expression (\ref{D2}), we get two explicit values for the roots of equation (\ref{discr}) with respect to $h_{n+2}$. Let us call them $x_+$ and $x_{-}$. By choosing the sign $+$ for the square root of (\ref{D2}) and by using the expression (\ref{Hn1}) and the differential equation (\ref{secor}) to eliminate the even powers of $\ddot{h}_n$ we get
\begin{equation}\label{fr}
x_{+}=h_n-\frac{ae_{n+1}}{2m\dot{h}_n}=h_n+\frac{e_{n+1}}{e_n}\left(h_{n+1}-h_{n-1}\right).
\end{equation}
Whereas, by choosing the sign $-$  for the square root of (\ref{D2}) and again by using the expression (\ref{Hn1}) and the differential equation (\ref{secor}) to eliminate the even powers of $\ddot{h}_n$ we get
\begin{equation}\label{tr}
x_{-}=h_n-\frac{4ae_{n+1}m\dot{h}_{n+1}^2}{2me_n\ddot{h}_n+2m^2(4a^2bh_n+g_2)\dot{h}_n+a^2e_n^2}.
\end{equation}
The root $x_{-}$ gives $h_{n+2}$. Indeed the expression (\ref{tr}) can be obtained by the repeated use of the B\"acklund transformation (\ref{Hn1}), i.e. by shifting the $n$ in (\ref{Hn1}) to $n+1$ and then using (\ref{Hn1}) again to get a formula relating $h_{n+2}$ to $h_n$. By doing this one obtains a huge formula for $h_{n+2}$ in terms of $h_n$ and its derivatives up to the fourth order, but   this formula can be simplified by taking into account the differential equation (\ref{secor}) and its differential consequences. What we get is exactly formula (\ref{tr}), i.e. $x_- = h_{n+2}$.

So we found the two roots of the quadratic recurrence (\ref{discr}): one is $h_{n+2}$, the other is given by $x_+$ in (\ref{fr}). At this point we can get explicit formulae relating $h_{n+2}$ to $h_{n+1}$, $h_n$ and $h_{n-1}$. Indeed, the two polynomials in $x$
\begin{equation}\label{discr1}
\begin{split}
&(x-h_{n+1})(x-h_n)(h_{n+1}-h_n)-\frac{m}{2ae_{n+1}}(x-h_{n})^2\left(4a^2bh_{n+1}+g_2\right)+\\
&-a^2(bz+c)\left(x-h_n\right)+\frac{ae_{n+1}}{2m}=0
\end{split}
\end{equation}
and
\begin{equation}\label{discr2}
\left(h_{n+1}-h_n-\frac{m(4a^2bh_{n+1}+g_2)}{2ae_{n+1}}\right)\left(x-h_n-\frac{e_{n+1}}{e_n}\left(h_{n+1}-h_{n-1}\right)\right)\left(x-h_{n+2}\right)
\end{equation}
must have the some roots. By equating the coefficients of $x$ and $x^0$ one gets two expressions for $h_{n+2}$ in terms of $h_{n+1}$, $h_n$ and $h_{n-1}$. These two expressions are indeed compatible with (\ref{h31}) and (\ref{h32}) if one takes into account equation (\ref{discr}) itself.

As far as we know both the discrete equation (\ref{discr}) satisfied by the Hamilton functions of the Painlev\'e equation II and the explicit solutions (\ref{h31}) and (\ref{h32}) are new. Once again, we stress that the first element of the chain $h_0$ could be a rational solution, a special type solution or a transcendental solution. The successive Hamiltonians can be obtained by means of (\ref{Hn1}) and satisfy (\ref{h31}), (\ref{h32}) and (\ref{discr})

In the next Section we consider further equation (\ref{linear}) and will give a linear equation satisfied by the corresponding tau functions.

\section{Linearization of the solutions of the Painlev\'e II equation}\label{sec4}
It is well known that special solutions of the Painlev\'e II equation can be obtained as solution of a Riccati equation. Indeed, it is the function $h_n(z)$ that satisfies a Riccati equation and the corresponding solutions of the Painlev\'e II are obtained through the relation (\ref{Hnn}). For completeness, we remember how the special solutions of Painlev\'e II are obtained. We look for solutions of equation (\ref{secor}) that are also solutions of a Riccati equation, i.e.
\begin{equation}\label{R1}
\dot{h}_n=Ah_n^2+Bh_n+C,
\end{equation}
where $A$, $B$ and $C$ are some indeterminate functions. By taking the derivative of (\ref{R1}), inserting it in (\ref{secor}) and taking into account (\ref{R1}) itself we get a sixth order polynomial for $h_n$. We look for conditions on the functions $A$, $B$ and $C$ so that all the coefficients of the powers of $h_n$ vanish. The coefficients of $h_n^6$ gives $A=-1$.   By inserting this value into the  sixth order polynomial we get a fourth order polynomial in $h_n$. The coefficient of $h_n^4$ is given by
\begin{equation}
C+\frac{B^2}{4}+\dot{B}+a^2(bz+c)
\end{equation}
and fixes the value of $C$ in terms of $B$. By inserting this into the fourth order polynomial we get a third order polynomial whose leading coefficient is given by $\ddot{B}$. So we set
\begin{equation}
B=\alpha+\beta z.
\end{equation}
With this value of $B$ we get a second order polynomial in $h_n$. The vanishing of the first and second order coefficients fix $\beta=0$ and $\alpha=-\frac{g_2}{2a^2b}$, whereas the order zero coefficients give the condition $e_n^2=4a^2b^2m^2$. By remembering that $e_n=e+2abmn$, it follows $e=0$ and $n=\pm 1$. So, if $e=0$ and $n=\pm 1$, equation (\ref{secor}) can be reduced to  
\begin{equation}
\dot{h}_n+h_n^2+\frac{g_2}{16b^2a^4}(8a^2bh_n+g_2)+a^2(bz+c)=0
\end{equation}
and, by letting $h_n=-\frac{g_2}{4a^2b}+\frac{\dot{\tau}_n}{\tau_n}$, one has
\begin{equation}
\ddot{\tau}_n=-a^2(bz+c)\tau_n,
\end{equation}
i.e. $\tau_n$ is an Airy function. Through the use of the B\"acklund transformations (\ref{Hn1}) and (\ref{invHn1}) one gets the family of special solutions corresponding to $e=0$ and $n \in \mathbb{N}$ in (\ref{secor}).

Now we will show that actually all the solutions of the equation (\ref{secor}) can be formally reduced to a solutions of a Riccati equation. The point is that the functions $A$, $B$ and $C$ in equation (\ref{R1}) could depend on a set of functions $h_{n+m}$ for some subset of integers $m$ and equation (\ref{R1}) could be satisfied due to some of the recurrences found in the above Sections. Let us start by noticing that, by evaluating (\ref{Hn1}) in $n \to n+1$ and subtracting with (\ref{invHn1}) it follows
\begin{equation}
h_{n+2}-h_n=-\frac{ae_{n+1}}{2m\dot{h}_{n+1}}.
\end{equation}
Differentiating we get
\begin{equation}\label{pro}
\dot{h}_{n+2}-\dot{h}_n=\frac{a e_{n+1}\ddot{h}_{n+1}}{2m(\dot{h}_{n+1})^2}.
\end{equation}
Now we look at the right hand side of (\ref{pro}). By equation (\ref{dhn}) it follows 
\begin{equation}\label{dhn1}
\dot{h}_{n+1}=-\frac{ae_{n+1}}{2m(h_{n+2}-h_{n})},
\end{equation}
and, from (\ref{invHn1}) and (\ref{dhn1}) we get
\begin{equation}\label{dd}
\frac{\ddot{h}_{n+1}}{\dot{h}_{n+1}}=h_{n+2}-2h_{n+1}+h_n.
\end{equation}
Equations (\ref{pro}), (\ref{dhn1}) and (\ref{dd}) together give
\begin{equation}\label{Ric}
\dot{h}_{n+2}=\dot{h}_n-\left(h_{n+2}-2h_{n+1}+h_n\right)\left(h_{n+2}-h_n\right).
\end{equation}
The previous is indeed a Riccati equation for $h_{n+2}$, with the coefficients $A$, $B$ and $C$ in (\ref{R1}) explicitly given by
\begin{equation}
A=-1, \quad B=2h_{n+1}, \quad C=\dot{h}_n+h_n^2-2h_nh_{n+1}.
\end{equation}
Notice that $h_{n+2}(z)$ is an arbitrary solution of equation (\ref{secor}) evaluated in $n \to n+2$, the constant $e_{n+2}$ being $e+2amb(n+2)$ with $e$ arbitrary.

Equation (\ref{Ric}) can be obtained also directly from (\ref{linear}) by evaluating it in $n \to n+1$ and subtracting (\ref{linear}) itself.

Let us introduce the $\tau_n$ functions through the logarithmic derivative
\begin{equation}\label{ht}
h_n(z)=\frac{\dot{\tau}_n}{\tau_n}.
\end{equation}
The functions $\tau_n$ are entire and possess simple zeros. Notice that the function $\tau_{n+2}$ solves a second order linear equation with coefficients depending on different $\tau$ functions. Explicitly we have
\begin{equation}
\tau_{n+1}\left(\ddot{\tau}_{n+2}\tau_n-\ddot{\tau}_n\tau_{n+2}\right)-2\dot{\tau}_{n+1}\left(\dot{\tau}_{n+2}\tau_n-\dot{\tau}_n\tau_{n+2}\right)=0.
\end{equation}
 The previous can be integrated once, giving
\begin{equation}\label{linear1}
\dot{\tau}_{n+2}\tau_n-\dot{\tau}_n\tau_{n+2}=\alpha_n\tau_{n+1}^2,
\end{equation}
where $\alpha_n$ is a constant that in principle could also depend on $n$. Notice that equation (\ref{linear1}) is linear in $\tau_n$ and $\tau_{n+2}$. Together with equation (\ref{linear1}) we should also consider the so called Toda equation, that can be obtained from (\ref{dd}). Indeed inserting definition (\ref{ht}) in (\ref{dd}) and integrating once we get
\begin{equation}\label{quad1}
\ddot{\tau}_{n+1}\tau_{n+1}-\dot{\tau}_{n+1}^2=\beta_n \tau_{n}\tau_{n+2},
\end{equation}
where again $\beta_n$ is a constant of integration. The constant $\alpha_n$ and $\beta_n$ actually depend on the gauge chosen to normalize each $\tau_n$ function: from (\ref{ht}) we see that indeed each $\tau_n$ is defined up to a scaling factor. We could fix it for example by taking the first non-zero Taylor series coefficients of $\tau_n$ in $z=0$ to be equal to 1. So, for example, if $z=0$ is not a zero of $\tau_n$ we can fix $\tau_n(z)=1+O(z)$ around $z=0$. However, from equation (\ref{dhn1}) we see that 
\begin{equation}\label{alb}
\alpha_n\beta_n=-\frac{ae_{n+1}}{2m}
\end{equation}
implying that, in general, they depend on $n$ and are both different from zero. Apart (\ref{linear1}) and (\ref{quad1}), other two differential equations can be obtained for the tau functions. They come directly from equations (\ref{dotHn}) and (\ref{linear}) and the use of (\ref{quad1}). We get:
\begin{equation}\label{eqn1}
\frac{a}{2m}\left(e_n\tau_n\dot{\tau}_{n+1}-e_{n+1}\tau_{n+1}\dot{\tau}_n\right)=\beta_n\beta_{n-1}\tau_{n+2}\tau_{n-1}+\frac{g_2}{4}\tau_n\tau_{n+1},
\end{equation}
and
\begin{equation}\label{eqn2}
\ddot{\tau}_{n+1}\tau_n+\ddot{\tau}_{n}\tau_{n+1}-2\dot{\tau}_{n+1}\dot{\tau}_{n}+a^2(bz+c)\tau_n\tau_{n+1}=0.
\end{equation}
This last equation can be cast also in the following form:
\begin{equation}
\ddot{y}_n+\left(\frac{\dot{\tau}_{n+1}}{\tau_{n+1}}-\frac{\dot{\tau}_n}{\tau_n}\right)^2+a^2(bz+c)=0, \quad y_n\doteq \ln(\tau_n\tau_{n+1})
\end{equation}
and, by introducing the  Wronskian between two successive tau functions, i.e.
\begin{equation}
W_n(z)=\dot{\tau}_{n+1}\tau_n-\tau_{n+1}\dot{\tau}_n = (h_{n+1}-h_{n})\tau_n\tau_{n+1},
\end{equation}
we see that it can be rewritten as
\begin{equation}
\ddot{y}_n+\left(\frac{W_{n}}{\tau_{n+1}\tau_n}\right)^2+a^2(bz+c)=0.
\end{equation}
Also, if we use (\ref{quad1}) we get:
\begin{equation}\label{Wsq}
\left(\frac{W_{n}}{\tau_{n+1}\tau_n}\right)^2+\frac{\beta_n\tau_n\tau_{n+2}}{\tau_{n+1}^2}+\frac{\beta_{n-1}\tau_{n-1}\tau_{n+1}}{\tau_{n}^2}+a^2(bz+c)=0.
\end{equation}
In the next Section two explicit formulae for $W_n$ in terms of the tau functions will be given.

\section{The ubiquitous role of the tau functions}\label{sec5}
In this Section we will show that all the functions appearing above, i.e. $q_n$, $p_n$ and $h_n$, can be expressed as rational functions of tau functions. Indeed, as enunciated before, it is possible to get an explicit formula for $W_n$ itself  (not its square) in terms of tau functions. To this aim, let us take the derivative of equation (\ref{linear1}) shifted in $n$ to $n-1$ and use equation (\ref{quad1}) and its suitable shift in $n$ to eliminate the second derivatives. We get:
\begin{equation}
\beta_n\frac{\tau_{n-1}\tau_n\tau_{n+2}}{\tau_{n+1}}-\beta_{n-2}\frac{\tau_{n-2}\tau_n\tau_{n+1}}{\tau_{n-1}}+\frac{\dot{\tau}_{n+1}^2\tau_{n-1}^2-\dot{\tau}_{n-1}^2\tau_{n+1}^2}{\tau_{n+1}\tau_{n-1}}=2\alpha_{n-1}\tau_n\dot{\tau}_n.
\end{equation}
We use again (\ref{linear1}) shifted in $n \to n-1$ in the previous to get:
\begin{equation}\label{eql}
\frac{\alpha_{n-1}\tau_n}{\tau_{n+1}\tau_{n-1}}\left(2\dot{\tau}_n\tau_{n+1}\tau_{n-1}-\dot{\tau}_{n+1}\tau_n\tau_{n-1}-\dot{\tau}_{n-1}\tau_n\tau_{n+1}\right)=\beta_n\frac{\tau_{n-1}\tau_n\tau_{n+2}}{\tau_{n+1}}-\beta_{n-2}\frac{\tau_{n-2}\tau_n\tau_{n+1}}{\tau_{n-1}},
\end{equation}
i.e., in terms of Wronskians
\begin{equation}\label{eqll}
\alpha_{n-1}\left(\tau_{n+1}W_{n-1}-\tau_{n-1}W_n\right)=\beta_n\tau_{n-1}^2\tau_{n+2}-\beta_{n-2}\tau_{n-2}\tau_{n+1}^2.
\end{equation}
Further, we notice that, due to equation (\ref{linear1}), it holds another linear relation between $W_{n}$ and its neighbours, i.e.
\begin{equation}\label{Wn}
\tau_{n+1}W_{n-1}+\tau_{n-1}W_{n}=\alpha_{n-1}\tau_n^3.
\end{equation}
Equations (\ref{Wn}) and (\ref{eqll}) can be solved for $W_n$ and $W_{n-1}$ giving 
\begin{align}
W_n=\frac{\alpha_{n-1}}{2}\frac{\tau_{n}^3}{\tau_{n-1}}+\frac{\beta_{n-2}\tau_{n+1}^2\tau_{n-2}-\beta_n\tau_{n-1}^2\tau_{n+2}}{2\alpha_{n-1}\tau_{n-1}}, \label{Wn1}\\
W_{n-1}=\frac{\alpha_{n-1}}{2}\frac{\tau_{n}^3}{\tau_{n+1}}-\frac{\beta_{n-2}\tau_{n+1}^2\tau_{n-2}-\beta_n\tau_{n-1}^2\tau_{n+2}}{2\alpha_{n-1}\tau_{n+1}}. \label{Wn2}
\end{align} 
The previous two equations are compatible with (\ref{Wsq}). Indeed, if we consider the difference between (\ref{Wsq}) evaluated in $n \to n+1$ and itself, then by using (\ref{Wn}) we arrive at formula (\ref{Wn1}) and (\ref{Wn2}) again.

Equations (\ref{Wn1}) and (\ref{Wn2}) are quite remarkable: indeed the left hand side are entire functions and so are the right hand side. It follows that the zeros of the denominator correspond to zeros of the numerator. We will research further in this direction in Section \ref{sec6}. Further, equations (\ref{Wn1}) and (\ref{Wn2}) give an explicit recursion for $\tau_{n+5}$ in terms of the four previous tau functions:
\begin{equation}\label{taurec}
\tau_{n+5}(z) = \frac{\alpha_{n+1}\alpha_{n+2} \tau_{n+2}(z) \tau_{n+4}(z) }{\beta_{n+3}\tau_{n+1}(z)}+\frac{ \alpha_{n+2} \beta_n\tau_n(z) \tau_{n+3}(z)^2 \tau_{n+4}(z)}{\alpha_{n+1} \beta_{n+3}\tau_{n+1}(z) \tau_{n+2}(z)^2 } + \frac{a^2b\tau_{n+1}(z)\tau_{n+4}(z)^2 }{\alpha_{n+1} \beta_{n+3}\tau_{n+2}(z)^2}-\frac{\alpha_{n+2}^2\tau_{n+3}(z)^3}{\beta_{n+3}\tau_{n+2}(z)^2 }  
\end{equation} 
Equation (\ref{taurec}), as far as we know, appeared first in \cite{JM} (see equation 5.7), but since then it seems to have been overlooked: it appeared also in \cite{Conte}, in \cite{FW} and, in relation to the rational solutions of Painlev\'e II, in \cite{Clarkson1}.

We end this Section by giving two equivalent expressions of the solution $q_n$ of the Painlev\'e equation II and of the Painlev\'e equation XXXIV as a rational combination of tau functions. From equations  (\ref{Hnn}), (\ref{Wn1}) and (\ref{Wn2}) we get
\begin{equation}\label{q1}
aq_n=-\frac{\alpha_{n-1}}{2}\frac{\tau_{n}^2}{\tau_{n+1}\tau_{n-1}}+\frac{\beta_{n-2}\tau_{n+1}^2\tau_{n-2}-\beta_n\tau_{n-1}^2\tau_{n+2}}{2\alpha_{n-1}\tau_{n-1}\tau_n\tau_{n+1}},
\end{equation}
and
\begin{equation}\label{q2}
aq_n=-\frac{\alpha_{n-2}}{2}\frac{\tau_{n-1}^2}{\tau_{n}\tau_{n-2}}+\frac{\beta_{n-1}\tau_{n-2}^2\tau_{n+1}-\beta_{n-3}\tau_{n}^2\tau_{n-3}}{2\alpha_{n-2}\tau_{n-2}\tau_{n-1}\tau_{n}}.
\end{equation}
The above equations are compatible through (\ref{taurec}). From the first equation in (\ref{tra}) we can get the expression for the solution of the Painlev\'e XXIV equation (\ref{eqmot}) $p_n$ in terms of tau functions. Since we have two different expressions for $q_n$ we have four different possibilities to write $q_n+q_{n+1}$. Of this four equations we write only the one being more compact, obtained by adding equation (\ref{q1}) with (\ref{q2}) evaluated at $n$ to $n+1$. We get
\begin{equation}\label{p1}
p_{n}=\frac{e_n\tau_{n-1}\tau_{n+1}}{\alpha_{n-1}\tau_{n}^2}=-\frac{2m}{a}\frac{\beta_{n-1}\tau_{n-1}\tau_{n+1}}{\tau_n^2}.
\end{equation} 
We can consider also the  second equation in (\ref{tra}): this will give a fourth order quadratic equation for the tau functions. Explicitly, from (\ref{tra}) and by using (\ref{p1}) and (\ref{q2}) we get
\begin{equation}\label{sectau}
\frac{1}{4}\left(\frac{\beta_{n+2}\tau_{n+1}\tau_{n+4}}{\alpha_{n+1}\tau_{n+2}\tau_{n+3}}+\frac{\alpha_{n+1}\tau_{n+2}^2}{\tau_{n+1}\tau_{n+3}}-\frac{\beta_{n}\tau_{n}\tau_{n+3}}{\alpha_{n+1}\tau_{n+1}\tau_{n+2}}\right)^2+\frac{\beta_n\tau_n\tau_{n+2}}{\tau_{n+1}^2}+\frac{\beta_{n+1}\tau_{n+1}\tau_{n+3}}{\tau_{n+2}^2}+a^2(bz+c)=0.
\end{equation}
Notice that if we used equation (\ref{q1}) instead of (\ref{q2}) in the second equation (\ref{tra}) we would have found the same quadratic equation for $\tau_{n+4}$ in terms of $\tau_{n+k}$, $k=0,...,3$. Obviously, equations (\ref{sectau}) and (\ref{taurec}) are compatible. Equation (\ref{sectau}) can be also obtained directly by using (\ref{Wsq}) and one of the two explicit equations for the Wronskian obtained above. The problem of the existence of an equation of lower degree for the $\tau$ functions (but maybe of higher degree) seems to be a difficult one to answer.

Finally, in this Section, we would like to present the explicit formula linking the Hamiltonian function $h_n$ to the tau functions.  From formula (\ref{HamH}), by remembering that $H_n=2bm h_n$, and from the above rational expressions of $q_n$ and $p_n$ in terms of tau functions it is possible to write different equivalent formulae for $h_n$. However, the corresponding equation is 
quite cumbersome. A more readable formula can be derived by considering the expressions for the derivative of a $\tau$ functions in terms of the tau functions themselves. Indeed, we notice that equation (\ref{eqn1}) and (\ref{Wn1}) are linear in the first derivative of $\tau_n$ and $\tau_{n+1}$, so, if $b\neq 0$, i.e. if we are not in the Weierstrass case, we can solve them to obtain:
\begin{equation}\label{der1}
\dot{\tau}_n = -\frac{\alpha_{n-1}^{2}\beta_{n-1}\tau_{n}^{3}}{2\alpha^{2}b\tau_{n+1}\tau_{n-1}} - \frac{\beta_{n-2}\beta_{n-1}\tau_{n+1}\tau_{n-2}}{2\alpha^{2}b\tau_{n-1}} - \frac{\beta_{n-1}\beta_{n}\tau_{n-1}\tau_{n+2}}{2\alpha^{2}b\tau_{n+1}} - \frac{g_{2}\tau_{n}}{4\alpha^{2}b},
\end{equation}
and
\begin{equation}\label{der2}
\dot{\tau}_{n+1}=\frac{\alpha_{n-2}\beta_{n-2}\beta_{n}\tau_{n-1}\tau_{n+2}}{2\alpha_{n-1}a^{2}b\tau_{n}} - \frac{g_{2}\tau_{n+1}}{4a^{2}b} - \frac{\alpha_{n-1}\alpha_{n}\beta_{n}\tau_{n}^{2}}{2a^{2}b\tau_{n-1}} - \frac{\alpha_{n}\beta_{n-2}\beta_{n}\tau_{n+1}^{2}\tau_{n-2}}{2\alpha_{n-1}a^{2}b\tau_{n}\tau_{n-1}}.
\end{equation}
So for example we get
\begin{equation}
h_n=-\frac{\alpha_{n-1}^{2}\beta_{n-1}\tau_{n}^{2}}{2\alpha^{2}b\tau_{n+1}\tau_{n-1}} - \frac{\beta_{n-2}\beta_{n-1}\tau_{n+1}\tau_{n-2}}{2\alpha^{2}b\tau_{n-1}\tau_n} - \frac{\beta_{n-1}\beta_{n}\tau_{n-1}\tau_{n+2}}{2\alpha^{2}b\tau_n\tau_{n+1}} - \frac{g_{2}}{4\alpha^{2}b}
\end{equation}
In the next Section we will look closer at the divisibility properties following from the expressions given in this Section relating the various tau functions with themselves.

\section{A Somos-4 like relation for the $\tau$ functions}\label{sec6}
The left hand side equations (\ref{der1}) and (\ref{der2}) are entire functions of $z$, so also the right hand side has to be an entire functions and the zeros of the denominators must be zeros of the corresponding numerators. In this Section we will present different relations of this type and we are going to exploit them to get some divisibility property between the two functions. So we are going to consider the divisibility properties of the tau functions in the ring of entire functions. To be precise, let us make the following
\begin{defn}\label{def}
Given two entire functions $f(z)$ and $g(z)$, we say that $g(z)$ divides $f(z)$ if there is another entire function of $z$, say $k(z)$, such that $f(z)=k(z)g(z)$, i.e. all the zeros of $g(z)$ are also zeros of $f(z)$. 
\end{defn}
It is well-known that $\tau_n$ and $\tau_{n+1}$ have no common zeros \cite{JM}. We notice also that, if $\alpha_n \neq 0$, from (\ref{linear1}) it follows that $\tau_n$ and $\tau_{n+2}$ cannot have common zeros (otherwise $\tau_n$ and $\tau_{n+1}$ would have a common zero). From the equations for the Wronskians (\ref{Wn1}) and (\ref{Wn2}) we get the following 
\begin{propn}
The function $\tau_n$ divides $\alpha_n^2\tau_{n+1}^3+\beta_{n-1}\tau_{n-1}\tau_{n+2}^2$, whereas the function $\tau_{n+2}$ divides $\alpha_n^2\tau_{n+1}^3+\beta_{n+1}\tau_n^2\tau_{n+3}$. More precisely one has
\begin{equation}\begin{split}
&\alpha_n^2\tau_{n+1}^3+\beta_{n+1}\tau_n^2\tau_{n+3}=\tau_{n+2}\left(2\alpha_nW_n+\beta_{n-1}\tau_{n-1}\tau_{n+2}\right)\\
&\alpha_n^2\tau_{n+1}^3+\beta_{n-1}\tau_{n-1}\tau_{n+2}^2=\tau_{n}\left(2\alpha_nW_{n+1}+\beta_{n+1}\tau_{n}\tau_{n+3}\right).
\end{split}\end{equation}
\end{propn} 
By taking the difference of the previous two equations and collecting $\tau_n$ and $\tau_{n+2}$ we have:
\begin{propn}\label{pro2}
The function $\tau_{n}$ divides $\beta_{n-1}\tau_{n-1}\tau_{n+2}+\alpha_{n}W_{n}$, whereas $\tau_{n+2}$ divides $\beta_{n+1}\tau_n\tau_{n+3}+\alpha_{n}W_{n+1}$, the quotient, say $x_n$, being the same.
\end{propn}
It is not difficult to show that the quotient is explicitly given by
\begin{equation}\label{xn}
x_n=\frac{\beta_{n-1}\tau_{n-1}\tau_{n+2}W_{n+1}-\beta_{n+1}\tau_{n}\tau_{n+3}W_n}{\tau_nW_{n+1}-\tau_{n+2}W_n}.
\end{equation}
Notice that in (\ref{xn}) also the zeros of the denominator are the zeros of the numerator. 

Other divisibility properties can be found by looking at the zeros of suitable expressions. Let us assume that $z_0$ is a zero of $\tau_{n+1}$. From Proposition (\ref{pro2}), since $\tau_{n}$ divides $\beta_{n-1}\tau_{n-1}\tau_{n+2}+\alpha_{n}W_{n}$, we have that in $z=z_0$ $\beta_{n}\tau_{n}\tau_{n+3}+\alpha_{n+1}W_{n+1}=0$. So we get an explicit expression for the derivative of $\tau_{n+1}$ evaluated in $z_0$:
\begin{equation}\label{r1}
\left.\frac{d\tau_{n+1}}{dz}\right|_{z=z_0}=\frac{\beta_n}{\alpha_{n+1}}\frac{\tau_{n+3}(z_0)\tau_n(z_0)}{\tau_{n+2}(z_0)}.
\end{equation}
Also, from equation (\ref{eqn1}) evaluated in $z=z_0$ we get
\begin{equation}\label{r2}
\left.\frac{d\tau_{n+1}}{dz}\right|_{z=z_0}=-\frac{\beta_n}{\alpha_{n-1}}\frac{\tau_{n+2}(z_0)\tau_{n-1}(z_0)}{\tau_{n}(z_0)}.
\end{equation}
From equation (\ref{quad1}) we have
\begin{equation}\label{r3}
\left(\left.\frac{d\tau_{n+1}}{dz}\right|_{z=z_0}\right)^2=-\beta_n\tau_n(z_0)\tau_{n+2}(z_0).
\end{equation}
Together, formula (\ref{r1}-\ref{r3}) gives
\begin{equation}
\alpha_{n-1}\alpha_{n+1}\tau_n(z_0)\tau_{n+2}(z_0)-\beta_{n}\tau_{n-1}(z_0)\tau_{n+3}(z_0)=0,
\end{equation}
where $z_0$ is any of the zeros of $\tau_{n+1}$. The previous combination of $\tau$ functions has the same set of zeros of $\tau_{n+1}$, so from Definition (\ref{def}) we get the following
\begin{propn}\label{propbo}
The function $\tau_{n+1}$ divides $\alpha_{n-1}\alpha_{n+1}\tau_n\tau_{n+2}-\beta_{n}\tau_{n-1}\tau_{n+3}$, i.e. it exists an entire function, say $y_{n+1}$, such that
\begin{equation}\label{eqpr}
\alpha_{n-1}\alpha_{n+1}\tau_n\tau_{n+2}-\beta_{n}\tau_{n-1}\tau_{n+3}=y_{n+1}\tau_{n+1}
\end{equation}
\end{propn} 
The relations (\ref{r1})-(\ref{r3}) actually define the values of $\alpha_n$ and $\beta_n$ in terms of the values of the tau functions evaluated in one of the zero of a specific tau functions. These relations can be used to get a more symmetric form of the equation (\ref{eqpr}). We can state Proposition (\ref{propbo}) in this alternative form:
\begin{propn}
Let $\omega$ be any of the zeros of $\tau_n$. Then the following equation holds:
\begin{equation}\label{eqpr1}
\frac{\tau_{n-2}(z)\tau_{n+2}(z)}{\tau_{n-2}(\omega)\tau_{n+2}(\omega)}-\frac{\tau_{n+1}(z)\tau_{n-1}(z)}{\tau_{n+1}(\omega)\tau_{n-1}(\omega)}=c_n(\omega)y_n(z)\tau_{n}(z),
\end{equation}
where $y_n(z)$ is given in (\ref{eqpr}) and $c_n(\omega)$ is given by
\begin{equation}
c_n(\omega)=\frac{\tau_{n-1}(\omega)\tau_{n+1}(\omega)}{\dot{\tau}_{n}(\omega)^2\tau_{n-2}(\omega)\tau_{n+2}(\omega)}
\end{equation}
\end{propn}
Clearly, from equation (\ref{eqpr1}), one has that the zeros of $y_n(z)$ are also the zeros of the left hand side. So if $\Omega$ is a zero of $y_n(z)$, we can write
\begin{equation}
\frac{\tau_{n-2}(\Omega)\tau_{n+2}(\Omega)}{\tau_{n-1}(\Omega)\tau_{n+1}(\Omega)}=\frac{\tau_{n-2}(\omega)\tau_{n+2}(\omega)}{\tau_{n-1}(\omega)\tau_{n+1}(\omega)}=\frac{\alpha_{n-2}\alpha_n}{\beta_{n-1}}
\end{equation}
Notice that, since when evaluated in $\omega$ (any of the zeros of $\tau_n(z)$) the values of $\tau_{n\pm1}(z)$ and $\tau_{n\pm2}(z)$ are different from zero, it follows that when evaluated in $\Omega$ (any of the zeros of $y_n(z)$) the values of $\tau_{n\pm1}(z)$ and $\tau_{n\pm2}(z)$ are different from zero. This may suggest that in some way the distributions of the zeros of $\tau_n(z)$ and of $y_n(z)$ are similar. We will show now that in the case $b=0$ the two functions are indeed proportional each other. In the case $b\neq 0$ we cannot give an analytical precise statement so far, but we will give some numerical observations. Before to consider the case $b=0$ let us make the following
\begin{rem}\label{remz} If $z_1$ is a zero of $\tau_n(z)$, it cannot be a zero of $\tau_{n\pm k}$ with $k=1,2$ if $\alpha_n \neq 0$. Suppose instead that $z_1$ is a common zero of $\tau_{n+2}$ and $\tau_{n-1}$ or a common zero of $\tau_{n-2}$ and $\tau_{n+1}$. From relation (\ref{eqpr1}) it follows that this zero must be a zero of $y_{n}$. 
\end{rem}
This Remark will be useful when we will inspect equation (\ref{eqpr}) in the case of polynomial solutions.  In the next subsections we will examine the Proposition (\ref{propbo}) starting from the subcase $b=0$ (Weierstrass case) and then to the polynomial solutions for $\tau_n$, the Airy-type solutions and finally the general transcendental case. These cases cover all the possible type of solutions of the Painlev\'e II equation (see e.g. \cite{UW}).

\subsection{The Weierstrass case}
Let us firstly notice that the function $y_n(z)$ can be expressed with a different combinations of $\tau$ functions. Indeed, in equation (\ref{eqpr}) we can use the explicit expression of $\tau_{n+3}$ in terms of $\tau_{n+k}$, $k=-2..2$ given by the shift of equation (\ref{taurec}). This give the following expression for $y_{n+1}(z)$:
\begin{equation}\label{sy}
\begin{split}
&y_{n+1}(z) = -\frac{\tau_{n+1}(z) \alpha_n \beta_n \left( -\tau_{n-1}(z) \tau_{n+1}(z) \alpha_n \alpha_{n-1} + \tau_{n-2}(z) \tau_{n+2}(z) \beta_{n-2} \right)}{\alpha_{n-1} \tau_n(z)^2 \beta_{n+1}} +\\
&- \frac{a^2b \tau_{n+2}(z) \left( \tau_n(z)^3 \alpha_{n-1}^2 + \tau_{n-1}(z)^2 \tau_{n+2}(z) \beta_n \right)}{\alpha_{n-1} \tau_n(z)^2 \beta_{n+1} \tau_{n+1}(z)}
\end{split}
\end{equation}
By using again the shift of (\ref{eqpr}) in $n\to n-1$ in the second addend of the right hand side of (\ref{sy}) for $\tau_{n+2}(z)$ and rearranging the terms we get
\begin{equation}\label{eqb}
\begin{split}
&y_{n+1}(z) \tau_n(z) \beta_{n-1} \alpha_{n-1} \beta_{n+1} - \tau_{n+1}(z) y_n(z) \alpha_n \beta_n \beta_{n-2} =\\
&= -\frac{a^2 b \left( \tau_{n-1}(z) \beta_n \alpha_n^2 \tau_{n+1}(z)^3 + \tau_{n+2}(z) \tau_n(z)^3 \alpha_{n-1}^2 \beta_{n-1} + \tau_{n-1}(z)^2 \tau_{n+2}(z)^2 \beta_n \beta_{n-1} \right)}{\tau_n(z) \tau_{n+1}(z)}.
\end{split}
\end{equation}
Equation (\ref{eqb}) simplifies drastically in the Weierstrass case $b=0$. Indeed one simply has
\begin{equation}\label{eqbW}
\frac{y_{n+1}(z)}{\tau_{n+1}(z)} - \frac{\beta_{n-2}}{\beta_{n+1}}\frac{y_n(z)}{\tau_{n}(z)}  =0,
\end{equation}
where we have taken into account that for $b=0$ the value of $\alpha_n\beta_n$ is independent of $n$ (see equation (\ref{alb})). We underline that with a proper choice of the gauge for each $\tau_n(z)$ it is possible to fix the value of $\beta_n$ to be a given constant, independent of $n$, say $\beta_n=\beta$. Since $\alpha_n\beta_n$ is also independent of $n$, with this choice of the gauge $\alpha_n$ is independent of $n$ too, say $\alpha_n=\alpha$. From equation (\ref{eqbW}) we see that $y_n=\tau_n$ (a constant in front of $\tau_n$ can be reabsorbed by redefining $\alpha$ and $\beta$) and equation (\ref{eqpr}) is, in this case, the Somos-4 relation:
\begin{equation}\label{eqwcase}
\alpha^2\tau_{n-1}(z)\tau_{n+1}(z)-\beta\tau_{n-2}(z)\tau_{n+2}(z)=\tau_n^2(z)
\end{equation}
It is well known that the general solution of this relation can be given in terms of the Weierstrass $\sigma$ functions (see e.g. \cite{H} and \cite{Swart}). From our point of view this can be seen also from relations (\ref{secor}). Indeed, for $b=c=0$, the general solution of (\ref{secor}) reads
\begin{equation}
h_n(z)=A_n+\zeta(z+B_n,g_g,-\alpha^2\beta^2),
\end{equation}
where $\zeta(z,g_2,g_3)$ is the Weierstrass zeta function with elliptic invariants $g_2$ and $g_3$  and $A_n$ and $B_n$ are the two constants of integration. It follows that $\tau_n(z)$ is given by
\begin{equation}\label{sigm}
\tau_n(z)=C_ne^{A_nz}\sigma(z+B_n,g_2,-\alpha^2\beta^2)
\end{equation}
where $\sigma(z,g_2,g_3)$ is the Weierstrass sigma function. The sequences $A_n$ and $B_n$ are fixed by the B\"acklund transformations (\ref{Hn1}) that now can be rewritten as
\begin{equation}\label{HnW}
h_{n+1}=h_n+\frac{\ddot{h}_n}{2\dot{h}_n}+\frac{\alpha\beta}{2\dot{h}_n}.
\end{equation}
By taking the first few terms of the Taylor series of (\ref{HnW}) around $z=-B_n$, one has
\begin{equation}\label{bcon}
A_{n+1}=A_n-\zeta(B_{n+1}-B_n,g_2,\alpha^2\beta^2), \quad \wp(B_{n+1}-B_n,g_2,\alpha^2\beta^2)=0, \quad \dot{\wp}(B_{n+1}-B_n,g_2,\alpha^2\beta^2)=\alpha\beta.
\end{equation}
The second equation in (\ref{bcon}) fixes the sequence $B_n$ as $B_n=B_0+nz_0$ where $z_0$ is a zero of $\wp(z,g_2,\alpha^2\beta^2)$. The first equation fixes the sequence $A_n$ as $A_n=A_0-n\zeta(z_0,g_2,\alpha^2\beta^2)$ whereas the third equation fixes the sign of $\dot{\wp}(z_0,g_2,\alpha^2\beta^2)$\footnote{Indeed, since $\wp(z_0,g_2\alpha^2\beta^2)=0$ one has $\dot{\wp}(z_0,g_2,\alpha^2\beta^2)^2=\alpha^2\beta^2$. The inverse transformation (\ref{invHn1}) gives the opposite sign for $\dot{\wp}(z_0,g_2,\alpha^2\beta^2)$.}. The values of the constants $C_n$ in (\ref{sigm}) must be chosen in such a way that $\beta_n=\beta$, i.e. does not depend on $n$. We need to look at the Toda equation (\ref{quad1}). We get
\begin{equation}\label{taubeta}
\frac{\ddot{\tau}_{n+1}\tau_{n+1}-\dot{\tau}_{n+1}^2}{\tau_n\tau_{n+2}}=-\frac{C_{n+1}^2}{C_nC_{n+2}}\frac{\sigma(z+B_{n+1})^2\wp(z+B_{n+1})}{\sigma(z+B_n)\sigma(z+B_{n+2})}=\beta 
\end{equation}
where we omit the invariants from now on. By setting $z+B_{n+1}=x$, using the well known formula \cite{WW} $\frac{\sigma(z+y)\sigma(z-y)}{\sigma^2(z)\sigma^2(y)}=\wp(y)-\wp(z)$ and by using $\wp(z_0)=0$ we get
 \begin{equation}
 -\frac{C_{n+1}^2}{C_nC_{n+2}}\frac{\sigma(z+B_{n+1})^2\wp(z+B_{n+1})}{\sigma(z+B_n)\sigma(z+B_{n+2})}= \frac{C_{n+1}^2}{C_nC_{n+2}\sigma(z_0)^2}=\beta. 
 \end{equation}
So the constants $C_n$ must be chosen in such a way that
\begin{equation}
\frac{C_{n+2}C_{n}}{C_{n+1}^2}=\frac{1}{\beta\sigma^2(z_0)},
\end{equation}
giving
\begin{equation}
C_n=\frac{C_1^nC_{0}^{1-n}}{\left(\beta^{1/2}\sigma(z_0)\right)^{n(n+1)}}
\end{equation}
and hence
\begin{equation}\label{hs}
\tau_n(z)=\hat{C}_0\hat{C}_1^n\frac{\sigma(z+B_0+nz_0)}{\left(\beta^{1/2}\sigma(z_0)\right)^{n^2}}
\end{equation} 
where we take into account the symmetry $\tau_n\to \tilde{A}\tilde{B}^n\tau_n$ for the equation (\ref{eqwcase}), i.e. if $\tau_n$ is a solution so is $\tilde{A}\tilde{B}^n\tau_n$. Notice that the factor $\beta^{1/2}$ can be removed from the denominator by a proper rescaling of $\alpha$ and $\beta$. Equation (\ref{hs}) can be found for example in \cite{H} and \cite{Swart}. In the Appendix we will consider the further degenerate case when the modular discriminant $g_2^3-27g_3^2$ vanishes, corresponding to trigonometric (or hyperbolic) solutions.

\subsection{The case of Yablonskii-Vorob'ev polynomials}
\begin{figure}[H]
\centering
\includegraphics[scale=0.7]{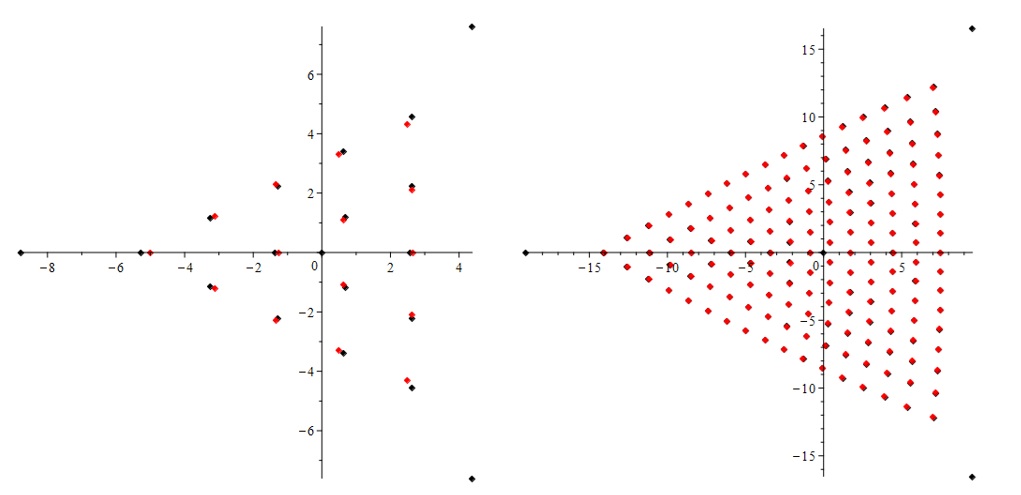}
\caption{On the left we plot of the real and imaginary part of the roots of $Q_5$ (red) and $P_5$ (black), on the right the roots of $Q_{17}$ (red) and $P_{17}$ (black).}
\label{fig1}
\end{figure}

Let us now consider the degenerate case when the functions $\tau_n(z)$ are proportional to the the Yablonskii-Vorob'ev polynomials $Q_n(z)$. In order to get these polynomials written in the standard form, one has to set $a=m=2b=1$, $c=g_2=0$, $e_n=n+1/2$. In the Toda equation (\ref{quad1}) one has to set $\beta_n=-1/4$ \cite{Taneda} (and hence $\alpha_n=2n+3$ from equation (\ref{alb})), whereas $\tau_n(z)$ is explicitly given by 
\begin{equation}\label{tq}
\tau_n(z)\doteq Q_n(z)e^{-\frac{z^3}{24}}, 
\end{equation}
where the recurrence solved by $Q_n$ follows from (\ref{quad1}) as  (see also \cite{Clarkson}, \cite{FOU}, \cite{Taneda}, \cite{V}, \cite{Y}):
\begin{equation}
Q_{n+2}=\frac{zQ_{n+1}^2-4\left(Q_{n+1}\ddot{Q}_{n+1}-\dot{Q}_{n+1}^2\right)}{Q_n}, \;\; Q_0 = 1, \; Q_1 = z.
\end{equation}
From the Proposition (\ref{propbo}) we get the 
\begin{cor}\label{corYV}
The Yablonskii-Vorob'ev polynomials $Q_{n}$ solve the following non-autonomous version of the Somos-4 relation
\begin{equation}\label{YVS}
4(2n-1)(2n+3)Q_{n-1}Q_{n+1}+Q_{n-2}Q_{n+2}=y_nQ_n=zP_nQ_n,
\end{equation}
where $P_n$ is a polynomial in $z$ of degree $3+\frac{n(n+1)}{2}$.
\end{cor}
The degree of $P_n$ follows directly from the fact that $Q_n$ has degree $\frac{n(n+1)}{2}$. The fact that $y_{n}(z)$ has always a root in $z=0$ follows from the properties of $Q_n$ and by considering (\ref{YVS}) modulo 3. Indeed, $Q_n$is divisible by $z$ if and only if $n=1\bmod 3$, $ Q_n$ is a polynomial in $z^3$ if $n\neq 1\bmod3$ and $Q_n(z)/z$ is a polynomial in $z^3$ if $n=1\bmod3$ \cite{Taneda}. By considering the left hand side of (\ref{YVS}), we see that it has a zero in $z=0$ for $n\neq 1\bmod3$, and this factor must be a zero of $y_n(z)$ (see Remark (\ref{remz})). For $n=1\bmod3$ we notice that the left hand side of (\ref{YVS}) is a polynomial in $z^3$, whereas the right hand side has a factor equal to $z$ from $Q_n$: it follows that $y_n$ has a factor equal to $z^2$ when $n=1\bmod3$.  The previous comments complete the Corollary (\ref{corYV}) with the following
\begin{propn}
The polynomial $P_n$ defined in (\ref{YVS}) is a polynomial in $z^3$ if $n\neq 1\bmod 3$, whereas $P_n/z$ is a polynomial in $z^3$ if $n=1\bmod 3$. 
\end{propn}
The first few $P_n$ are listed below
\begin{equation*}\begin{aligned}
P_0 &= z^3 - 8 \\
P_1 &= z(z^3 + 40) \\
P_2 &= z^6 + 140z^3 + 1120 \\
P_3 &= z^9 + 300z^6 + 8400z^3 - 22400 \\
P_4 &= z(z^{12} + 532z^9 + 36960z^6 + 172480z^3 + 6899200) \\
P_5 &= z^{18} + 852z^{15} + 122640z^{12} + 3523520z^9 + 76876800z^6 - 2690688000z^3 - 7175168000 \\
P_6 &= z^{24} + 1280z^{21} + 338800z^{18} + 26364800z^{15} + 948640000z^{12} - 13812198400z^9 + 1736390656000z^6 \\
&\quad + 27624396800000z^3 - 44199034880000
\end{aligned}\end{equation*}
For comparison, we list here also the first few Yablonskii-Vorob'ev polynomials:
\begin{equation*}
\begin{aligned}
Q_0 &= 1 \\
Q_1 &= z \\
Q_2 &= z^3 + 4 \\
Q_3 &= z^6 + 20z^3 - 80 \\
Q_4 &= (z^9 + 60z^6 + 11200)z \\
Q_5 &= z^{15} + 140z^{12} + 2800z^9 + 78400z^6 - 3136000z^3 - 6272000 \\
Q_6 &= z^{21} + 280z^{18} + 18480z^{15} + 627200z^{12} - 17248000z^9 + 1448832000z^6\\
     &\quad  + 19317760000z^3 - 38635520000 
\end{aligned}
\end{equation*}
In the next figure we plot the roots of some $Q_n$ and $P_n$. Numerically we notice that one can associate at a root of $Q_n$ the closest root of $P_n$. As $n$ increases the distance between corresponding roots decreases. We do not have, at the moment, any quantitative statement about this phenomenon. Also, since $Q_n$ is of degree $\frac{n(n+1)}{2}$ and $P_n$ of degree $\frac{n(n+1)}{2}+3$, there are three roots of $P_n$ not associated to any roots of $Q_n$. These three roots, say given numerically by $z^{*}$, $\omega z^*$ and $\omega^2 z^{*}$, where $\omega$ is the cubic root of unity, are, in modulus, greater than the other roots and so, as can be seen also from Figure (\ref{fig1}), define a triangle that contains all the other roots. For other aspects of the rational solutions to Painlev\'e II, like for example asymptotic behaviour in $z$, the reader can look for example in \cite{Miller} and references therein.

\subsection{The Airy-type solutions}
Now we consider the Airy-type solutions for Painlev\'e II. The parameters can fixed as follows (see e.g. \cite{Clarkson}). The value of $\beta_n$ can be fixed to be $-1$, then the value of $\alpha_n$ is fixed from (\ref{alb}) as $\alpha_n=a^2b(n+1)$, since one has to set $e=0$ in the definition of the values of $e_n=e+2abmn$. Similar to the rational case, the tau functions have the following structure: the same exponential factor, given by $e^{-\frac{g_2}{4a^2b}z}$ times a polynomial of Airy functions and their derivatives. We remember equation (\ref{kn}) and set:
\begin{equation}\label{tauairy}
\tau_n(z)=e^{-\frac{g_2}{4a^2b}z} \Psi_n(z), \quad h_{n}(z)=-\frac{g_2}{4a^2b} + k_n(z),
\end{equation}
where the relation between $\Psi_n$ and $k_n$ is 
\begin{equation}
k_n=\frac{\dot{\Psi}_n}{\Psi_n}.
\end{equation}  
It is easily see that the first two Airy polynomials are given by
\begin{equation}\label{P0P1}
\Psi_0(z)=1, \quad \Psi_1(z)\doteq \ddot{\Psi}_1+a^2(bz+c)\Psi_1=0.
\end{equation}
The function $\Psi_1$ can be explicitly written as a linear combination of $\textrm{Ai}\left(-\frac{(a^2b)^{1/3}(bz+c)}{b}\right)$ and $\textrm{Bi}\left(-\frac{(a^2b)^{1/3}(bz+c)}{b}\right)$, where Ai$(z)$ is, up to a constant factor, the only solution of the Airy differential equation $\ddot{y}=zy$ exponentially decaying on the positive real axis. The function Bi$(z)$ is instead exponentially increasing on the positive real axis (see e.g. \cite{Z2, Z1}). The other Airy-type solutions follow from the recursion
\begin{equation}\label{recaity}
\Psi_{n+2}=\frac{-\Psi_{n+1}\ddot{\Psi}_{n+1}+\dot{\Psi}_{n+1}^2}{\Psi_n}.
\end{equation}
Each $\Psi_n$ is a polynomial in $\Psi_1$ and $\dot{\Psi}_1$.

From Proposition (\ref{propbo}) we get the following
\begin{cor}
The Airy-type solution $\Psi_n$, polynomial of $\Psi_1$ and $\dot{\Psi}_1$, solves the following non-autonomous Somos-4 type relation for $n\geq 2$
\begin{equation}\label{SomosAiry}
a^4b^2(n^2-1)\Psi_{n-1}\Psi_{n+1}+\Psi_{n-2}\Psi_{n+2}=y_n\Psi_n,
\end{equation}
where $y_n$ is another polynomial of $\Psi_1$ and $\dot{\Psi}_1$.
\end{cor} 
Actually, the polynomials defining $\Psi_n$ and $y_n$ have some nice algebraic properties. For example we have the following
\begin{propn} The polynomials $\Psi_n$ and $y_n$ are homogeneous of degree $n$ in $\Psi_1$ and $\dot{\Psi}_1$, i.e.  $\Psi_n=\sum_{k=0}^n c_k\dot{\Psi}_1^k\Psi_1^{n-k}$ and $y_n=\sum_{k=0}^n \hat{c}_k\dot{\Psi}_1^k\Psi_1^{n-k}$, for some $c_k$ and $\hat{c}_k$ polynomials in $z$. 
\end{propn}
Indeed, by using $\ddot{\psi}_1=-a^2(bz+c)\psi_1$, it is easy to show that if $\Psi_n$ is an homogeneous polynomial of degree $n$ in $\Psi_1$ and $\dot{\Psi}_1$, so are $\dot{\Psi}_n$ and $\ddot{\Psi}_n$. By induction from (\ref{recaity}) then it follows that $\Psi_n$ is homogeneous of degree $n$ for all $n$. So we have that $\Psi_n(\lambda \Psi_1, \lambda \dot{\Psi}_1)=\lambda^n\Psi_n(\Psi_1,\dot{\Psi}_1)$. Applying this relation to the recursion (\ref{SomosAiry}), we get that also $y_n$ is a homogeneous polynomial of degree $n$ in $\Psi_1$ and $\dot{\Psi}_1$.

Here we list the first few Airy-type solutions (see also \cite{Clarkson}) and the first few Airy-type functions $y_n$. For brevity in this list we define $A(z)=a^2(bz+c)$, so that $\psi_1$ is any solution of $\ddot{\psi}_1=-A\psi_1$. 

\begin{equation*}
\begin{aligned}
\Psi_2&=\dot{\Psi}_1^2 + A \Psi_1^2 \\
\Psi_3&=- 2 \dot{A} \left( \dot{\Psi}_1^3 + A \Psi_1^2 \dot{\Psi}_1 - \frac{1}{2} \dot{A} \Psi_1^3 \right) \\
\Psi_4&=4 \dot{A}^2 \left( A \dot{\Psi}_1^4 + 2 \dot{A} \Psi_1 \dot{\Psi}_1^3 + 2 A^2 \Psi_1^2 \dot{\Psi}_1^2 + 2 \dot{A} A \Psi_1^3 \dot{\Psi}_1 + \Psi_1^4 \left( A^3 - \frac{3}{4} \dot{A}^2 \right) \right)\\
\Psi_5&=-32 \dot{A}^{4} \left( \frac{9}{4} \dot{A} \dot{\Psi}_{1}^{5} + A^{2} \Psi_{1} \dot{\Psi}_{1}^{4} + \frac{11}{2} A \Psi_{1}^{2} \dot{A} \dot{\Psi}_{1}^{3} + 2 \Psi_{1}^{3} \left( A^{3} - \frac{7}{16} \dot{A}^{2} \right) \dot{\Psi}_{1}^{2} + \frac{13}{4} A^{2} \Psi_{1}^{4} \dot{A} \dot{\Psi}_{1} + A \Psi_{1}^{5} \left( A^{3} - \frac{9}{8} \dot{A}^{2} \right) \right)\\
y_{2} &= 4 \dot{A}^{2} \left( A \dot{\Psi}_{1}^{2} + \frac{1}{2} \dot{A} \Psi_{1} \dot{\Psi}_{1} + A^{2} \Psi_{1}^{2} \right) \\
y_{3} &= -16 \dot{A}^{3} \left( A \dot{\Psi}_{1}^{3} - \frac{1}{4} \dot{A} \Psi_{1} \dot{\Psi}_{1}^{2} + A^{2} \Psi_{1}^{2} \dot{\Psi}_{1} - \frac{3}{4} \dot{A} A \Psi_{1}^{3} \right) \\
y_{4} &= 64 \dot{A}^{4} \left( A^{2} \dot{\Psi}_{1}^{4} + \frac{7}{4} \dot{A} A \Psi_{1} \dot{\Psi}_{1}^{3} + 2 \Psi_{1}^{2} \left( A^{3} + \frac{\dot{A}^{2}}{32} \right) \dot{\Psi}_{1}^{2} + \frac{7}{4} \dot{A} A^{2} \Psi_{1}^{3} \dot{\Psi}_{1} + A \Psi_{1}^{4} \left( A^{3} - \frac{9}{16} \dot{A}^{2} \right) \right) \\
 y_{5} &= -768 \dot{A}^{6} \left( \frac{5}{2} \dot{A} A \dot{\Psi}_{1}^{5} + \Psi_{1} \left( A^{3} + \frac{5 \dot{A}^{2}}{16} \right) \dot{\Psi}_{1}^{4} + 6 \dot{A} A^{2} \Psi_{1}^{2} \dot{\Psi}_{1}^{3} + 2 A \Psi_{1}^{3} \left( A^{3} - \frac{5 \dot{A}^{2}}{16} \right) \dot{\Psi}_{1}^{2} + \right.\\
 &\left.+\frac{7}{2} \dot{A} \Psi_{1}^{4} \left( A^{3} - \frac{\dot{A}^{2}}{28} \right)+ \dot{\Psi}_{1} + A^{2} \Psi_{1}^{5} \left( A^{3} - \frac{19 \dot{A}^{2}}{16} \right) \right)
\end{aligned}
\end{equation*}
For other aspects of the Airy type solutions, like for example asymptotic behaviour in $z$, the reader can look for example in (\ref{D1}) and references therein.
\subsection{The general case}
Finally, we consider the general transcendental case. Let us fix the normalization of the tau functions just in the same way as we did for the previous cases, i.e. we choose $\beta_n=\beta$ independent of $n$. Each function $\tau_n(z)$ solves a third order differential equation with respect to $z$ (see also \cite{HZ1}): this comes from the second order differential equation solved by $h_n$ (\ref{secor}) and $\frac{\dot{\tau}_n}{\tau_n}=h_n$. This differential equation reads

\begin{equation}\label{taun3}
\begin{split}
&\dddot{\tau}_{n}^{2} \tau_{n}^{2} - 6 \dddot{\tau}_{n}\ddot{\tau}_{n}\dot{\tau}_{n} \tau_{n} + 4 \dddot{\tau}_{n} \dot{\tau}_{n}^{3} + 4\ddot{\tau}_{n}^{3}\tau_{n} -\left( 3 \dot{\tau}_{n}^{2} - 4 a^{2} (bz + c) \tau_{n}^{2} \right) \ddot{\tau}_{n}^{2} +\\
& -\left( 8 a^{2} (bz + c) \tau_{n} \dot{\tau}_{n}^{2} + 4  a^{2} b\tau_{n}^{2} \dot{\tau}_{n} + g_2\tau_{n}^{3} \right) \ddot{\tau}_{n} + 4 a^{2} (bz + c) \dot{\tau}_{n}^{4} + 4 a^{2} b\tau_{n} \dot{\tau}_{n}^{3}  +g_2\tau_{n}^{2} \dot{\tau}_{n}^{2}  - \alpha_{n-1}^{2} \beta_{n-1}^{2}\tau_{n}^{4} =0
\end{split}
\end{equation}
Suppose to fix $\tau_0(z)$, a solution of (\ref{taun3}) for $n=0$. The equation solved by $\tau_1(z)$ comes from the B\"acklund transformation (\ref{Hn1}) and reads\footnote{We are assuming that we are in the general case, so the denominator of the right hand side of (\ref{tau1def}) is well defined. In the Airy case, when the general equation (\ref{tau1def}) does not apply, one has to use (\ref{P0P1}) instead of (\ref{tau1def}).}:
\begin{equation}\label{tau1def}
\dot{\tau}_{1} = \frac{\left( \alpha_{-1} \beta \tau_{0}^{2} - \dot{\tau}_{0} \ddot{\tau}_{0} + \dddot{\tau_0} \tau_{0} \right) \tau_{1}}{2 \left( \ddot{\tau}_{0} \tau_{0} - \dot{\tau}_{0}^{2} \right)}
\end{equation} 
The other tau functions can be recursively be defined with
\begin{equation}\label{recaity1}
\tau_{n+2}=\frac{\tau_{n+1}\ddot{\tau}_{n+1}-\dot{\tau}_{n+1}^2}{\beta\tau_n}.
\end{equation}
From equation (\ref{alb}) we see that if $\beta_n=\beta$ then $\alpha_n=\frac{e+2abm(n+1)}{\beta}$, where $\frac{e}{2abm}\notin \mathbb{Z}$ since $\alpha_n\neq 0$ for all $n$. From Proposition (\ref{propbo}) we get
\begin{cor}
The transcendental functions of $\tau_n$ defined by (\ref{tau1def}) and (\ref{recaity1}), where $\tau_0$ is a transcendental solution of (\ref{taun3}) with $\beta_0=\beta$ and $\alpha_0=\frac{e+2abm}{\beta}$, solve the following non-autonomous Somos-4 type relation
\begin{equation}
\left((e+2abmn)^2-(2abm)^2\right)\tau_{n-1}\tau_{n+1}-\beta^3\tau_{n-2}\tau_{n+2}=\beta^2y_n\tau_n
\end{equation}
where $y_n$ are transcendental entire functions.
\end{cor}

\subsection{Wronskian relations}
In this subsection we will show that there are different other equations that can, in principle, give some direction in the research of other divisibility properties for the tau functions. Let us define the Wronskian between $\tau_{n+k}$ and $\tau_n$ as $W_{n}^{(k)}$. Then the following factorization holds
\begin{equation}
W_{n}^{(k+2)}\tau_{n+2}=W_{n}^{(2)}\tau_{n+k+2}+W_{n+2}^{(k)}\tau_n.
\end{equation}
The previous is just an algebraic relation that can be easily verified, but with the use of $W^{(1)}_n=W_n$, where $W_n$ is given by one of the two formulae (\ref{Wn1}) and (\ref{Wn2}), and of $W_n^{(2)}=\alpha_n\tau_{n+1}^2$ gives a chain of equivalences solved by the tau functions. These equivalences define divisibility properties. The first two expressions that we find are

\begin{equation}
\begin{aligned}
&\dot{\tau}_{n+3}\tau_n-\dot{\tau}_n\tau_{n+3}= \frac{\alpha_{n+1}^2 \tau_{n+2}^3 \tau_n + 2\alpha_n \alpha_{n+1} \tau_{n+1}^3 \tau_{n+3} + \beta_n \tau_{n+3}^2 \tau_n^2 - \beta_{n+2} \tau_n \tau_{n+1}^2 \tau_{n+4}}{2\alpha_{n+1} \tau_{n+1} \tau_{n+2}}\\
&\dot{\tau}_{n+4}\tau_n-\dot{\tau}_n\tau_{n+4}=\frac{\alpha_{n+2}\tau_{n+3}^2\tau_n+\alpha_n\tau_{n+4}\tau_{n+1}^2}{\tau_{n+2}}
\end{aligned}
\end{equation}
Again, our observation is that the left hand side is an entire function, so must be the right hand side.

\section{Conclusions} The generalized structures presented in this work suggest that the solutions of the Painlev\'e II equation, Painlev\'e XXXIV equation, the Yablonskii-Vorob'ev polynomials, the Airy type solutions of the Painlev\'e II, the Weierstrass elliptic functions and the Hamiltonians of the Painlev\'e II equation can be considered the members of an extended  Painlev\'e II family of functions. All the discrete equations solved by these functions can be interpreted in terms of the fifth order difference equation (\ref{taurec}). Also, we have shown that the tau functions solve a Somos-4 like relation. There are different points not yet discussed in this  paper and that will be the subject of future works:  from the interlacing of the poles and zeros of the various functions, to some properties of the Yablonskii-Vorobev that follow for example from equation (\ref{linear1}). How much of these structures can be generalized to the entire Painlev\'e hierarchy, especially the Somos-4 type relations, is an open question. The starting point surely can be the corresponding B\"acklund transformations (see e.g. \cite{CHJ}).  It would interesting also to investigate possible connections with other aspects and properties of the tau functions, like their Fredholm determinant representation (see e.g. \cite{Des}). Finally, we believe that the structures here described could be lifted-up at the Painlev\'e equation IV: this also will be the subject of future research.

\section*{Appendix: the degenerate trigonometric case.}
It can be instructive to see what are the explicit expressions of the functions $p_n$, $q_n$, $h_n$ and $\tau_n$ in the trigonometric degenerate case, i.e. when the solutions of equation (\ref{secor}) can be written in terms of trigonometric functions. Actually, this corresponds to the Weierstrass reduction $b=0$ with the further constraint on the discriminant of the elliptic curve  $\Delta=g_2^3-27g_3^2=0$. Now the value of $e_n$ is fixed since $b=0$. The values of the constants $g_2$ and $e_n$ in (\ref{secor}) must be fixed to be equal to the following quantities:
\begin{equation}\label{he}
g_2=\frac{4}{3}\left(k^4-a^4c^2\right), \quad \left(\frac{ae_n}{2m}\right)^2=-\frac{4}{27}a^6c^3+\frac{4}{9}a^2ck^4-\frac{8}{27}k^6.
 \end{equation}
where $k$ is an arbitrary parameter. The corresponding equation for $h_n$ now reads
\begin{equation}\label{hnw0}
(\ddot{h}_n)^2 + 4 \frac{\left(3\dot{h}_n+a^2c+2k^2\right)\left(3\dot{h}_n+a^2c-k^2\right)^2}{27}=0.
\end{equation}
It is not difficult to show that the general solution of the previous equation is given by
\begin{equation}\label{hnw}
h_n=A_n+k\cot(k(z+B_n))+\frac{k^2-a^2c}{3}z
\end{equation}
where $A_n$ and $B_n$ are two arbitrary constants. The B\"acklund transformations define a dynamic on the constants $A_n$ and $B_n$. It is useful to parametrize the constant $c$ itself in terms of a trigonometric function
\begin{equation}
c=-\frac{k^2}{a}\left(2+3\cot(kw)^2\right).
\end{equation}  
With this parametrization, equation (\ref{hnw0}) takes the form
\begin{equation}
(\ddot{h}_n)^2 + 4 \left(\dot{h}_n-k^2\cot(kw)^2\right)\left(\dot{h}_n-k^2-k^2\cot(kw)^2\right)^2,
\end{equation}
with $h_n=A_n+k\cot(k(z+B_n))+\frac{k^2}{\sin(kw)^2}z$. The choice of $e_n$ in (\ref{he}) corresponds to the choice of forward or backward transformations. In terms of trigonometric functions, the forward transformation (\ref{Hn1}) corresponds to
\begin{equation}
\frac{e_na}{2m}=\frac{2k^3\cos(kw)}{\sin(kw)^3}. 
\end{equation}
The B\"acklund transformations define the following dynamics on $A_n$ and $B_n$:
\begin{equation}
A_{n+1}=A_n-k\cot(w(B_{n+1}-B_n)), \quad \cot(k(B_{n+1}-B_n))^2=\cot(kw)^2,
\end{equation}
that can be solved as
\begin{equation}
A_n=A_0-nk\cot(kw), \quad B_n=B_0+nw.
\end{equation}
Again, the arbitrariness of the sign in front of $w$ on the right hand side of the formula for $B_n$ corresponds to the choice of the forward or backward transformation.  The function $p_n$ is given by
\begin{equation}
p_n=\frac{2mk^2}{a}\left(\cot(k(z+B_n))^2-\cot(kw)^2\right),
\end{equation}
whereas for $\tau_n$ one has
\begin{equation}
\tau_n=\frac{\sin(k(z+B_n))}{\sin(kB_n)}\exp\left(A_nz+\frac{k^2z^2}{2\sin^2(kw)}\right)
\end{equation}
where we assumed that $B_n\neq 0$ and the normalization for $\tau_n$ is such that $\tau_n=1+O(z)$ for $z\to 0$. If, for some $j\in \mathbb{Z}$, $B_j=0$ (i.e. $B_0=-jw$), then one has
\begin{equation}
\tau_j=\frac{\sin(kz)}{k}\exp\left(A_jz+\frac{k^2z^2}{2\sin^2(kw)}\right),
\end{equation}
the corresponding normalization being given by $\tau_j(z)=z+O(z)^2$. The constants $\alpha_n$ and $\beta_n$ are given by
\begin{equation}
\alpha_n=-2k\frac{\cot(kw)\sin(kB_{n+1})^2}{\sin(kB_n)\sin(kB_{n+2})}, \quad \beta_n=k^2\frac{\sin(kB_n)\sin(kB_{n+2})}{\sin(kw)^2\sin(kB_{n+1})^2}
\end{equation}

\begin{center} {\bf Acknowledgments} \end{center}
The Authors wish to acknowledge the support of Universit\`a degli Studi di Brescia and of GNFM-INdAM. F. Z. wishes to acknowledge also INFN, Gr. IV - Mathematical Methods in Nonlinear Physics and ISNMP - International Society of Nonlinear Mathematical Physics.


\begin{thebibliography}{10}

\bibitem{Clarkson1}
P.~A. Clarkson, \emph{Remarks on the {Y}ablonskii-{V}orob'ev polynomials},
  Phys. Lett. A \textbf{319} (2003), no.~1-2, 137--144.

\bibitem{Clarkson}
\bysame, \emph{Painlev\'e{} equations---nonlinear special functions},
  Orthogonal polynomials and special functions, Lecture Notes in Math., vol.
  1883, Springer, Berlin, 2006, pp.~331--411.

\bibitem{C1}
\bysame, \emph{Open problems for {P}ainlev\'e{} equations}, SIGMA Symmetry
  Integrability Geom. Methods Appl. \textbf{15} (2019), Paper No. 006, 20.

\bibitem{CHJ}
P.~A. Clarkson, A.~N.~W. Hone, and N.~Joshi, \emph{Hierarchies of difference
  equations and {B}\"acklund transformations}, J. Nonlinear Math. Phys.
  \textbf{10} (2003), 13--26.

\bibitem{Des}
H.~Desiraju, \emph{Fredholm determinant representation of the homogeneous
  {P}ainlev\'e{} {II} {$\tau$}-function}, Nonlinearity \textbf{34} (2021),
  no.~9, 6507--6538.

\bibitem{FGR}
A.~S. Fokas, B.~Grammaticos, and A.~Ramani, \emph{From continuous to discrete
  {P}ainlev\'e{} equations}, J. Math. Anal. Appl. \textbf{180} (1993), no.~2,
  342--360.

\bibitem{FW}
P.~J. Forrester and N.~S. Witte, \emph{Application of the {$\tau$}-function
  theory of {P}ainlev\'e{} equations to random matrices: {PIV}, {PII} and the
  {GUE}}, Comm. Math. Phys. \textbf{219} (2001), no.~2, 357--398.

\bibitem{FOU}
S.~Fukutani, K.~Okamoto, and H.~Umemura, \emph{Special polynomials and the
  {H}irota bilinear relations of the second and the fourth {P}ainlev\'e{}
  equations}, Nagoya Math. J. \textbf{159} (2000), 179--200.

\bibitem{GLS}
V.~I. Gromak, I.~Laine, and S.~Shimomura, \emph{Painlev\'e{} differential
  equations in the complex plane}, De Gruyter Studies in Mathematics, vol.~28,
  Walter de Gruyter \& Co., Berlin, 2002.

\bibitem{H}
A.~N.~W. Hone, \emph{Elliptic curves and quadratic recurrence sequences}, Bull.
  London Math. Soc. \textbf{37} (2005), no.~2, 161--171, Corrigendum
  \textbf{38} (2006), no. 5, 741--742.

\bibitem{HZ2}
A.~N.~W. Hone, O.~Ragnisco, and F.~Zullo, \emph{Properties of the series
  solution for {P}ainlev\'e{} {I}}, J. Nonlinear Math. Phys. \textbf{20}
  (2013), 85--100.

\bibitem{HZ1}
A.~N.~W. Hone and F.~Zullo, \emph{A {H}irota bilinear equation for
  {P}ainlev\'e{} transcendents {$P_{IV}$}, {$P_{II}$} and {$P_I$}}, Random
  Matrices Theory Appl. \textbf{7} (2018), no.~4, 1840001, 15.

\bibitem{Ince}
E.~L. Ince, \emph{Ordinary {D}ifferential {E}quations}, Dover Publications, New
  York, 1944.

\bibitem{JM}
M.~Jimbo and T.~Miwa, \emph{Monodromy preserving deformation of linear ordinary
  differential equations with rational coefficients. {II}}, Phys. D \textbf{2}
  (1981), no.~3, 407--448.

\bibitem{Miller}
P.~D. Miller and Y.~Sheng, \emph{Rational solutions of the {P}ainlev\'e-{II}
  equation revisited}, SIGMA Symmetry Integrability Geom. Methods Appl.
  \textbf{13} (2017), Paper No. 065, 29.

\bibitem{Okamoto1}
K.~Okamoto, \emph{On the {$\tau $}-function of the {P}ainlev\'e{} equations},
  Phys. D \textbf{2} (1981), no.~3, 525--535.

\bibitem{Okamoto2}
\bysame, \emph{Studies on the {P}ainlev\'e{} equations. {III}. {S}econd and
  fourth {P}ainlev\'e{} equations, {$P_{{\rm II}}$} and {$P_{{\rm IV}}$}},
  Math. Ann. \textbf{275} (1986), no.~2, 221--255.

\bibitem{Conte}
\bysame, \emph{The {H}amiltonians associated to the {P}ainlev\'e{} equations},
  The {P}ainlev\'e{} property, CRM Ser. Math. Phys., Springer, New York, 1999,
  pp.~735--787.

\bibitem{Swart}
C.~S. Swart, \emph{Elliptic curves and related sequences}, Phd thesis, Royal
  Holloway, , University of London, 2003.

\bibitem{Taneda}
M.~Taneda, \emph{Remarks on the {Y}ablonskii-{V}orob'ev polynomials}, Nagoya
  Math. J. \textbf{159} (2000), 87--111.

\bibitem{UW}
H.~Umemura and H.~Watanabe, \emph{Solutions of the second and fourth
  {P}ainlev\'e{} equations. {I}}, Nagoya Math. J. \textbf{148} (1997),
  151--198.

\bibitem{V}
A.~P. Vorob'ev, \emph{On the rational solutions of the second {P}ainlev\'e{}
  equation}, Differencial'nye Uravnenija \textbf{1} (1965), 79--81.

\bibitem{WW}
E.~T. Whittaker and G.~N. Watson, \emph{A course of modern analysis. {A}n
  introduction to the general theory of infinite processes and of analytic
  functions: with an account of the principal transcendental functions}, fourth
  ed., Cambridge University Press, New York, 1962, Unaltered reprinting of the
  fourth edition [Cambridge Univ. Press, Cambridge, 1927].

\bibitem{Y}
A.~I. Yablonski\u{\i}, \emph{On rational solutions of the second painlev\'e
  equation}, Vesci Akad. Navuk BSSR Ser. Fiz.-T\`ehn. Navuk \textbf{3} (1959),
  30--35.

\bibitem{Z0}
F.~Zullo, \emph{B\"acklund transformations and {H}amiltonian flows}, J. Phys. A
  \textbf{46} (2013), no.~14, 145203, 16.

\bibitem{Z2}
\bysame, \emph{On the dynamics of the zeros of solutions of the {A}iry
  equation}, Math. Comput. Simulation \textbf{176} (2020), 312--318.

\bibitem{Z1}
\bysame, \emph{On the solutions of the {A}iry equation and their zeros},
  Complex differential and difference equations, De Gruyter Proc. Math., De
  Gruyter, Berlin, 2020, pp.~267--281.

\end{thebibliography}
%
%

\providecommand{\bysame}{\leavevmode\hbox to3em{\hrulefill}\thinspace}
\providecommand{\MR}{\relax\ifhmode\unskip\space\fi MR }
\providecommand{\MRhref}[2]{%
  \href{http://www.ams.org/mathscinet-getitem?mr=#1}{#2}
}
\providecommand{\href}[2]{#2}

\end{document}